\documentclass{aa}
\usepackage{psfig}
\usepackage{times}
\def\mag{\hbox{$^{\rm m}$}}
\def\degr{$^{\circ}$}
\begin{document}

\thesaurus{06               
           (02.19.1;        
            08.03.4;        
            08.06.2;        
            09.08.1;        
            09.10.1;        
            13.09.6)}       

\title{ISO Spectroscopy of Young Intermediate-Mass Stars in the BD+40\degr4124 Group
\thanks{Based on observations with ISO, an ESA project with instruments
        funded by ESA Member States (especially the PI countries: France,
        Germany, the Netherlands and the United Kingdom) and with the
        participation of ISAS and NASA.}}

\author{M.E. van den Ancker\inst{1,2} \and P.R. Wesselius\inst{3} \and 
        A.G.G.M. Tielens\inst{3,4,5}}
\institute{
Astronomical Institute ``Anton Pannekoek'', University of Amsterdam, 
 Kruislaan 403, NL--1098 SJ  Amsterdam, The Netherlands \and
Harvard-Smithsonian Center for Astrophysics, 60 Garden Street, MS 42, 
 Cambridge, MA 02138, USA \and 
SRON, P.O. Box 800, NL--9700 AV  Groningen, The Netherlands \and
Kapteyn Astronomical Institute, Groningen University, P.O. Box 800, 
 NL--9700 AV  Groningen, The Netherlands \and
NASA Ames Research Center, MS 245-3, Moffett Field, CA 94035, USA}

\offprints{M.E. van den Ancker (mario@astro.uva.nl)}
\date{Received <date>; accepted <date>} 

\maketitle

\begin{abstract}
We present the results of ISO SWS and LWS grating scans towards the 
three brightest members of the BD+40\degr4124 group in the infrared: 
BD+40\degr4124 (B2Ve), LkH$\alpha$ 224 (A7e) and the embedded 
source LkH$\alpha$ 225. Emission from the pure rotational lines 
of H$_2$, from ro-vibrational transitions of CO, 
from PAHs, from H\,{\sc i} recombination lines and from 
the infrared fine structure lines of 
[Fe\,{\sc ii}], [Si\,{\sc ii}], [S\,{\sc i}], [O\,{\sc i}], 
[O\,{\sc iii}] and [C\,{\sc ii}] was detected. These 
emission lines arise in the combination of a low-density 
($\approx$ 10$^2$~cm$^{-3}$) H\,{\sc ii} region with 
a clumpy PDR in the case of BD+40\degr4124. 
The lower transitions of the infrared H\,{\sc i} lines observed 
in BD+40\degr4124 are optically thick; most likely they arise 
in either a dense wind or a circumstellar disk. This same region 
is also responsible for the optical H\,{\sc i} lines and the 
radio continuum emission. In the lines of sight towards 
LkH$\alpha$ 224 and LkH$\alpha$ 225, the observed emission 
lines arise in a non-dissociative shock produced by a slow 
($\approx$ 20~km~s$^{-1}$) outflow arising from LkH$\alpha$ 225. 
Toward LkH$\alpha$ 225 we also observe a dissociative 
shock, presumably located closer to the outflow source than the 
non-dissociative shock. In the line of sight towards LkH$\alpha$ 225 
we observed absorption features due to solid water ice and amorphous 
silicates, and due to gas-phase H$_2$O, CO and CO$_2$. No solid CO$_2$ 
was detected towards LkH$\alpha$ 225, making this the first line of 
sight where the bulk of the CO$_2$ is in the gas-phase.
\keywords{Shock waves -- Circumstellar matter -- Stars: Formation -- 
          H\,{\sc ii} regions -- ISM: Jets and outflows-- Infrared: Stars}

\end{abstract}

\section{Introduction}
\begin{figure*}
\centerline{\psfig{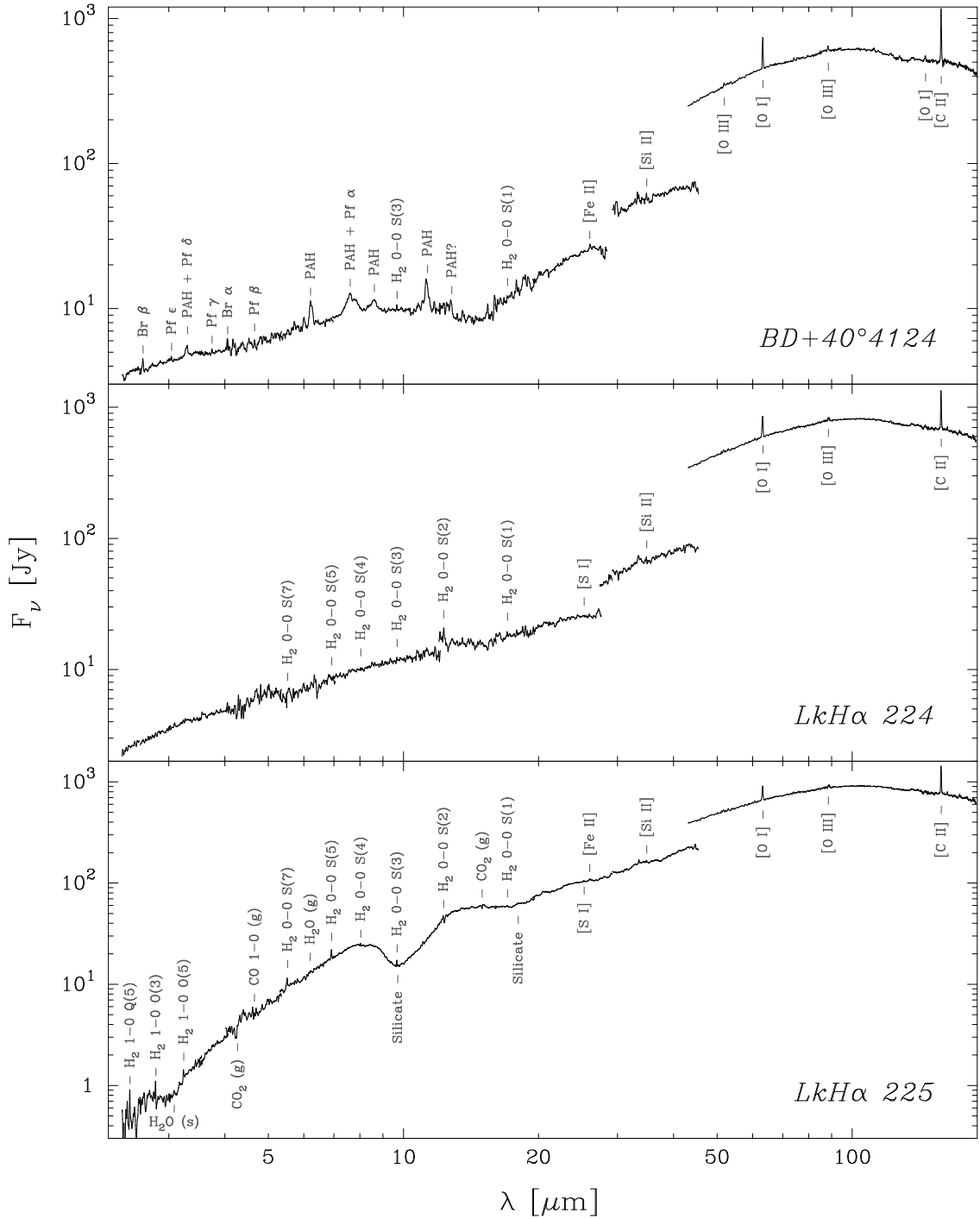}}
\caption[]{Combined SWS/LWS full grating spectra of (from top to bottom) 
BD+40\degr4124, LkH$\alpha$ 224 and LkH$\alpha$ 225 with the most 
prominent features identified. Apparent discontinuities in the 
spectra at 12.1, 27.5, 29.2 and 45.5~$\mu$m are caused by the 
change to a bigger aperture size and may reflect the presence 
of extended far-infrared emission. 
}
\end{figure*}
The BD+40\degr4124 region consists of a few tens of very young 
stars associated with the probable Herbig Ae/Be stars BD+40\degr4124 
(= V1685~Cyg), LkH$\alpha$ 224 (= V1686~Cyg) and the young multiple 
system LkH$\alpha$~225 
(= V1318~Cyg), all of which have infrared excesses (Strom et al. 1972a; 
Hillenbrand et al. 1995). Together with NGC 6910, NGC 6914, 
the BD+41\degr3731 region and IC 1318, it is part of the giant star 
forming region 2 Cyg, at a distance of about 1~kpc (Shevchenko 
et al. 1991). The stars in the BD+40\degr4124 region are 
significantly younger (ages less than 1 million years) than those 
in the surrounding OB associations, with the low- and the high-mass 
stars having formed nearly simultaneously, leading some authors 
to suggest that star formation in this association might have been 
induced by the propagation of an external shock wave into the 
cloud core (Shevchenko et al. 1991; Hillenbrand et al. 1995). 
The IRAS and AFGL surveys showed a powerful infrared source within 
the BD+40\degr4124 group, but the positional uncertainty did not 
allow to assign the flux to one of the individual objects in the 
region. More recent submm continuum maps of the region (Aspin et al. 
1994; Di Francesco et al. 1997; Henning et al. 1998) clearly 
peak on LkH$\alpha$ 225, suggesting that it is also dominant in 
the far-infrared and may be identified with IRAS 20187+4111 
(AFGL 2557).

Aspin et al. (1994) have shown LkH$\alpha$~225 to be a triple system
oriented north-south. The most northern and southern components, 
separated by 5\arcsec, appear stellar and are photometrically variable 
at optical and near-infrared wavelengths. The middle component 
exhibits strong [S\,{\sc ii}] emission, suggesting it is a nebulous 
knot of shock-excited material, not unlike many Herbig-Haro objects 
(Magakyan \& Movseyan 1997). Whereas BD+40\degr4124 dominates 
the association in the optical, the
southern component of LkH$\alpha$~225 is brightest in the mid-infrared 
(Aspin et al. 1994). A K-band spectrum of LkH$\alpha$~225-South by 
the same authors shows strong ro-vibrational emission lines of 
molecular hydrogen. Aspin et al. (1994) estimate the total luminosity 
of this object to be $\approx$ 1600~L$_\odot$, which places 
it in the luminosity range of the Herbig Ae/Be stars.

Palla et al. (1994) obtained near infrared, CO, C$^{18}$O, CS, and
H$_2$O maser observations of the group. Their high resolution VLA
data show that a H$_2$O maser source is clearly associated with
LkH$\alpha$~225-South. Moreover, a density concentration in the molecular
cloud (as evidenced by CS~J=5--4 emission) and a CO outflow are both
associated with LkH$\alpha$~225. In their model, LkH$\alpha$~225 is at the
center of a dense molecular core of mass $\approx$ 280~M$_\odot$, while
BD+40\degr4124 lies near the periphery. Continuum radio emission has 
been detected from BD+40\degr4124 and a source 42\arcsec~to the east 
of BD+40\degr4124 (Skinner et al. 1993).

The infrared emission-line spectrum of young stellar objects (YSOs) such 
as those in the BD+40\degr4124 group is dominated by the interaction 
of the central object with the remnants of the cloud from which it 
formed. If a young stellar object produces intense UV radiation, either 
due to accretion or because of a high temperature of the central star, 
an H\,{\sc ii} region will develop, producing a rich ionic emission 
line spectrum. At the edge of the H\,{\sc ii} region, a so-called 
photodissociation or photon-dominated region (PDR) will be created, 
in which neutral gas is heated by photoelectrically ejected electrons 
from grain surfaces. Cooling of this neutral gas occurs mainly through 
emission in atomic fine-structure and molecular lines, allowing us 
to probe the physical conditions in the PDR through the study of its 
infrared emission-line spectrum. If the YSO has a strong, often 
collimated, outflow, it will cause a shock wave as the outflow penetrates 
in the surrounding molecular cloud, heating the gas. In case the shock 
has a velocity smaller than $\approx$ 40~km~s$^{-1}$, it will not 
dissociate the molecular material and mainly cool through 
infrared molecular transitions. Because the physical conditions 
from the pre-shock gas to the post-shock gas change gradually 
within such a non-dissociative shock, they are often called 
C(ontinuous)-shocks. For higher shock velocities, the 
molecules will be dissociated and cooling in the shock occurs mainly 
through atomic and ionic emission lines. In such a shock, the 
physical conditions from the pre- to the post-shock gas will change 
within one mean free path and therefore they are usually termed 
J(ump)-shocks. In the post-shock gas of a J-shock, 
molecules will re-form and cool down the gas further. Therefore 
a dissociative shock will produce an infrared spectrum consisting 
of a combination of ionic and molecular emission lines.

In this paper we present {\it Infrared Space Observatory} (ISO; 
Kessler et al. 1996) spectroscopic data obtained at the positions of 
BD+40\degr4124, LkH$\alpha$ 224 and LkH$\alpha$ 225. Some of the H$_2$ 
lines included in this data-set were already analyzed in a previous 
{\em letter} (Wesselius et al. 1996). Here we will re-analyze an extended 
set of H$_2$ data using the latest ISO calibrations, in combination 
with new data on infrared fine-structure lines, gas-phase molecular 
absorption bands and solid-state emission features in the BD+40\degr4124 
group. We will show that the infrared emission lines arise from the 
combination of a H\,{\sc ii} region and a PDR at the position of 
BD+40\degr4124, from a non-dissociative shock at the position of 
LkH$\alpha$ 224, and from the combination of a 
non-dissociative and a dissociative shock at the 
position of LkH$\alpha$ 225. We will also discuss the unusual 
absorption-line spectrum seen in the line of sight towards 
LkH$\alpha$ 225 and briefly discuss its implications for the 
physical and chemical evolution of hot cores.

\section{Observations}
We obtained ISO Short Wavelength (2.4--45~$\mu$m) Spectrometer 
(SWS; de Graauw et al. 1996a) and Long Wavelength (43--197~$\mu$m) 
Spectrometer (LWS; Clegg et al. 1996) grating scans at 
the positions of BD+40\degr4124, LkH$\alpha$ 224 and LkH$\alpha$ 225. 
For all three positions, both SWS scans covering the entire SWS 
wavelength range (``AOT 01'') and much deeper scans covering small 
wavelength regions around particularly useful diagnostic lines 
(``AOT 02'') were obtained. For LWS only scans covering the entire 
wavelength region were obtained. A full log of the observations 
is given in Table~1.
\begin{figure*}
\centerline{\psfig{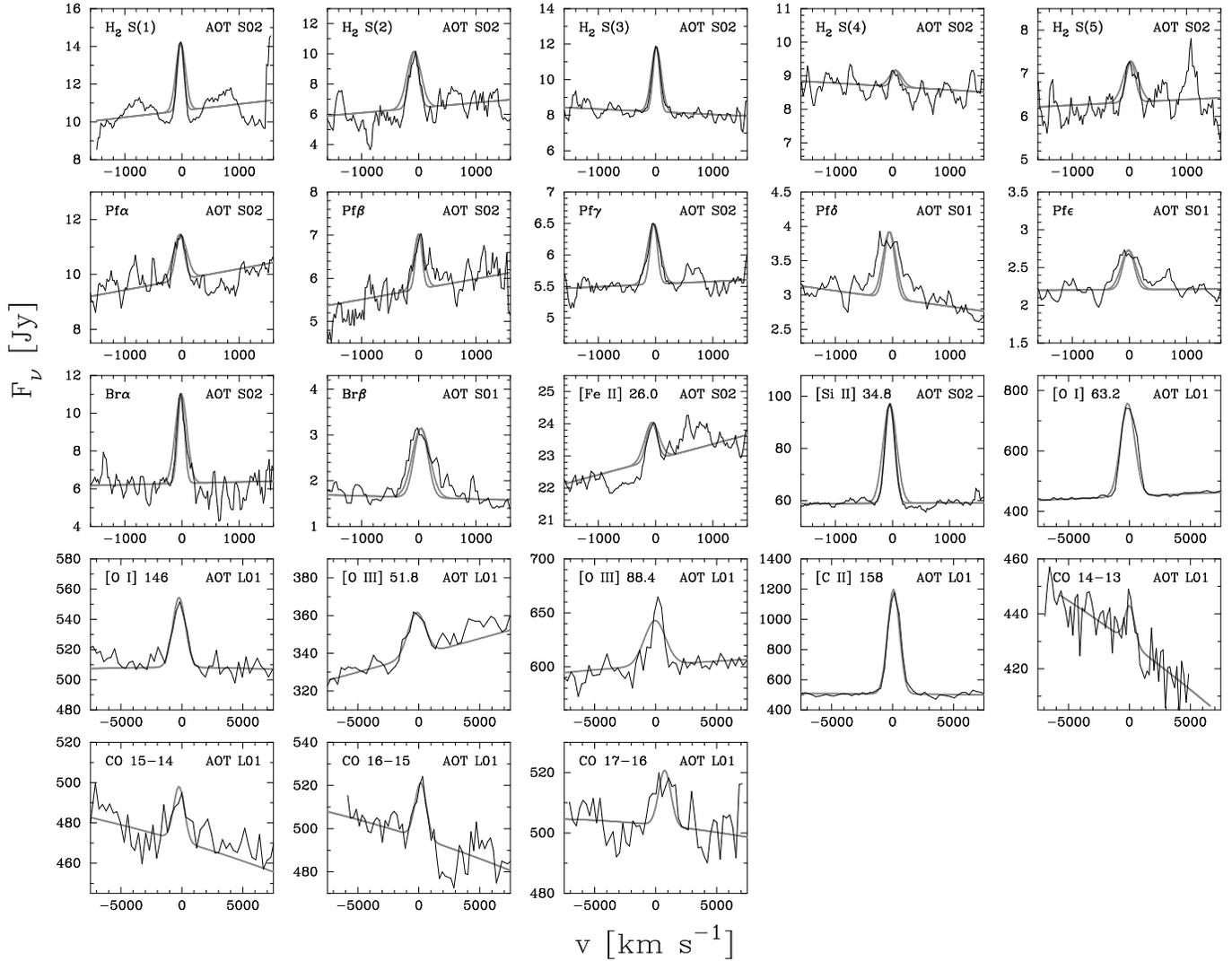}}
\caption[]{Detected lines in BD+40\degr4124, rebinned to heliocentric 
velocities. The grey lines show the instrumental profiles for a 
point source and an extended source filling the entire aperture.}
\end{figure*}
\begin{figure*}
\centerline{\psfig{figure=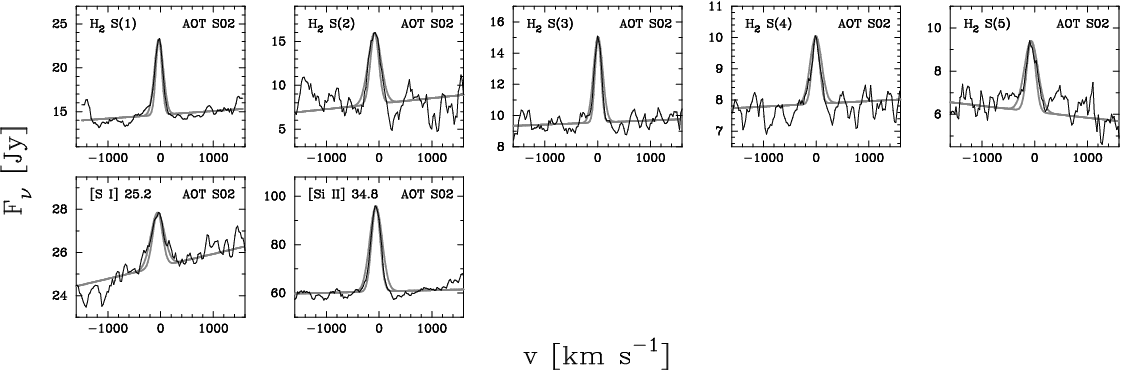,width=18.0cm}}
\caption[]{Same as Fig.~2 for LkH$\alpha$ 224.}
\end{figure*}
\begin{figure*}
\centerline{\psfig{figure=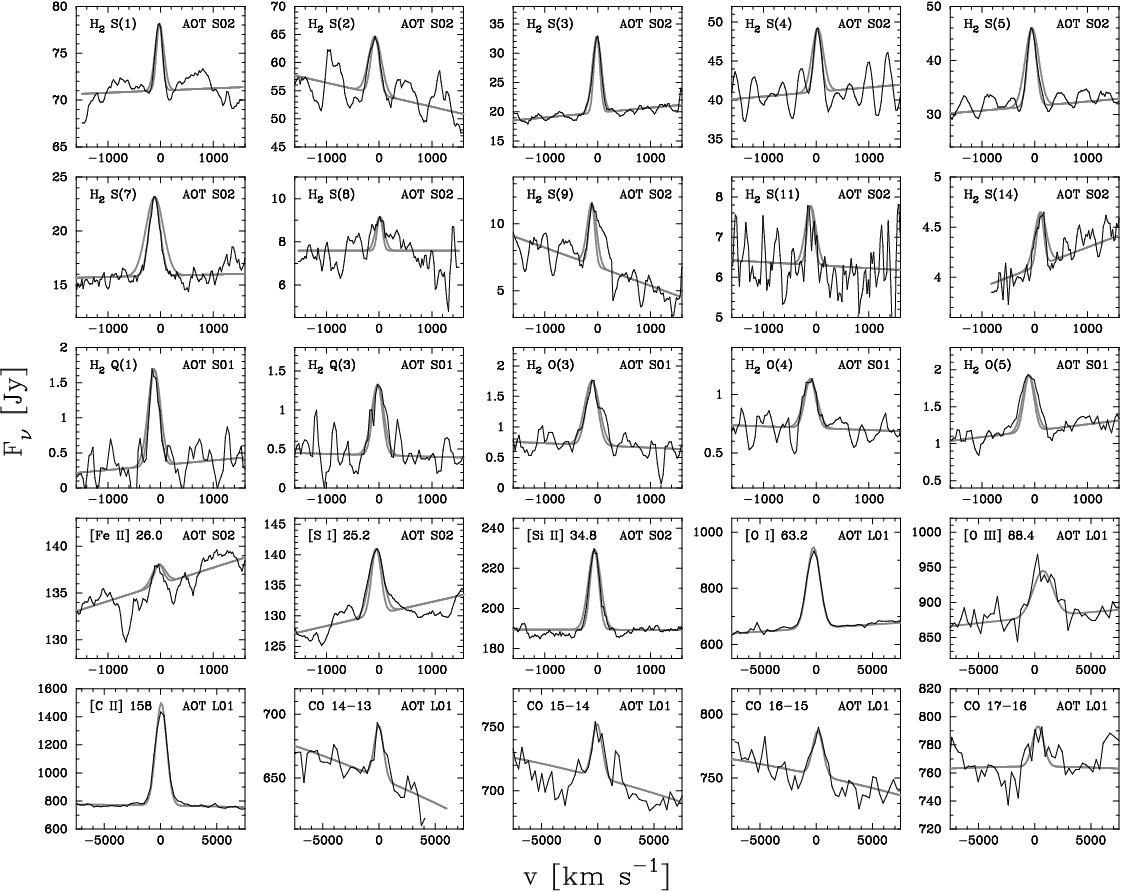,width=18.0cm}}
\caption[]{Same as Fig.~2 for LkH$\alpha$ 225.}
\end{figure*}
\begin{table}
\caption[]{Log of ISO observations in the BD+40\degr4124 region.}
\begin{flushleft}
\begin{tabular}{lcccc}
\hline\noalign{\smallskip}
Object         & AOT     & Rev. & Date         & JD$-$2450000\\
\noalign{\smallskip}
\hline\noalign{\smallskip}
BD+40\degr4124  & SWS 02 & 142  & 07/04/1996 & 181.037\\
                & SWS 02 & 159  & 24/04/1996 & 197.635\\
                & SWS 01 & 355  & 05/11/1996 & 393.283\\
                & LWS 01 & 768  & 23/12/1997 & 805.525\\
LkH$\alpha$ 224 & SWS 02 & 142  & 07/04/1996 & 180.960\\
                & SWS 01 & 858  & 22/03/1998 & 894.920\\
LkH$\alpha$ 225 & SWS 02 & 142  & 07/04/1996 & 180.996\\
                & SWS 02 & 355  & 05/11/1996 & 393.306\\
                & SWS 01 & 858  & 22/03/1998 & 894.889\\
                & LWS 01 & 142  & 07/04/1996 & 181.017\\
\noalign{\smallskip}
\hline
\end{tabular}
\end{flushleft}
\end{table}

Data were reduced in a standard fashion using calibration files 
corresponding to ISO off-line processing software (OLP) version 7.0, after 
which they were corrected for remaining fringing and glitches. To increase 
the S/N in the final spectra, the detectors were aligned and statistical 
outliers were removed, after which the spectra were rebinned to a lower 
spectral resolution. Figure~1 shows the resulting ISO spectra. 
Plots of all detected lines, rebinned to a resolution 
$\lambda/\Delta\lambda$ of 2000 with an oversampling factor of four 
(SWS), or averaged across scans (LWS), are presented in Figs.~2--4.
Line fluxes for detected lines and upper limits (total flux for line 
with peak flux 3$\sigma$) for the most significant undetected lines are 
listed in Table~2. 

For BD+40\degr4124 and LkH$\alpha$ 224, all detected lines 
appear at their rest wavelength within the accuracy of the ISO 
spectrometers. For LkH$\alpha$ 225, all lines in the SWS spectra 
are systematically shifted by $\approx$ $-$100~km~s$^{-1}$. This 
could either be a reflection of the real spatial velocity of 
LkH$\alpha$ 225 or be caused by a slight offset between our 
pointing, based on the optical position of the source, and the 
position of the infrared source. For all three sources, the 
lines are unresolved, indicating that the velocity dispersion 
is smaller than a few hundred km~s$^{-1}$.

For each complete spectral scan, the SWS actually makes twelve different 
grating scans, each covering a small wavelength region (``SWS band''), 
and with its own optical path.
They are joined to form one single spectrum (Fig.~1). Because of the 
variation of the diffraction limit of the telescope with wavelength, 
different SWS bands use apertures of different sizes. For a source 
that is not point-like, one may therefore see a discontinuity in flux  
at the wavelengths where such a change in aperture occurs. This effect 
can indeed be seen in some of our spectra, indicating the presence of 
extended far-infrared emission throughout the region.

Since both grating spectrometers on board ISO use apertures that are 
fairly large compared to the separation of sources in most star forming 
regions, some caution is appropriate in interpreting such measurements. 
We created a plot with the positions of the SWS apertures, overlaid on 
a K$^{\prime}$-band image of the region (Hillenbrand et al. 1995), 
shown in Fig.~5. As can be seen from this figure, only one strong 
near-infrared source is included in the BD+40\degr4124 and 
LkH$\alpha$ 224 measurements, whereas the LkH$\alpha$ 225 measurement 
only includes the question-mark shaped infrared nebula formed by 
the point-like objects LkH$\alpha$ 225-North and LkH$\alpha$ 225-South 
and the bridging nebula LkH$\alpha$ 225-Middle. 
ISO-LWS has a beam that is larger (61--83\arcsec~FWHM) than 
the separation of the objects studied here. Therefore the LWS 
measurements of BD+40\degr4124 and LkH$\alpha$ 225 partly cover 
the same region. The LWS spectrum of LkH$\alpha$ 224 
shown in Fig.~1 is a weighted average of these spectra.

Accuracies of the absolute flux calibration in the SWS spectra range 
from 7\% in the short-wavelength ($<$ 4.10~$\mu$m) part to $\approx$ 
30\% in the long wavelength ($>$ 29~$\mu$m) part (Leech et al. 1997). 
The LWS absolute flux calibration is expected to be accurate at the 
7\% level (Trams et al. 1997). Note that the applied flux calibration 
is based on the assumption that we are looking at a point source. 
For an extended source, the diffraction losses will be underestimated 
and in particular at the long-wavelength part of the LWS, these 
corrections will exceed the quoted uncertainty.
%
\begin{figure*}
\vspace*{-6cm}
\hspace*{-2cm}
\centerline{\psfig{figure=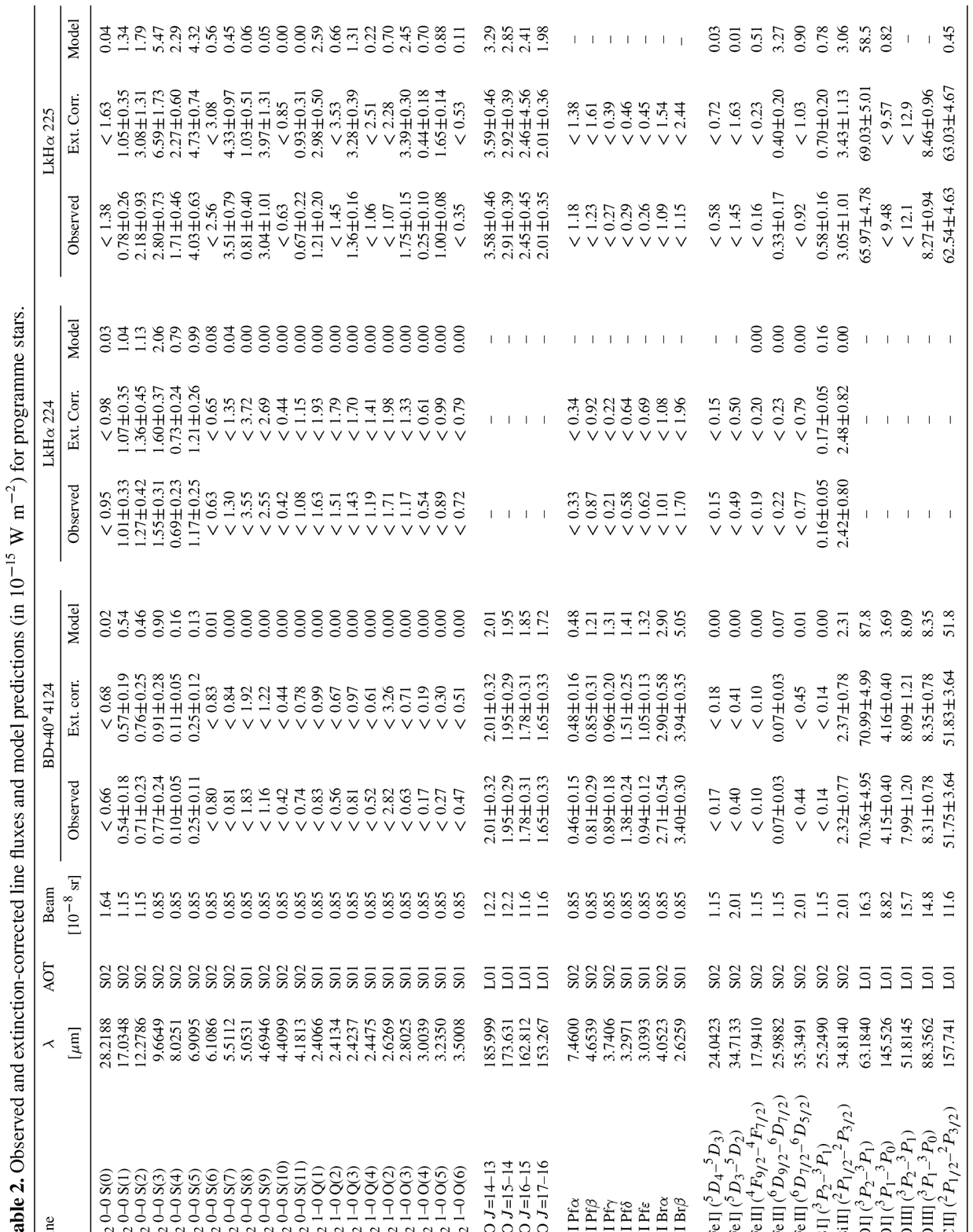,angle=180}}
\end{figure*}
\addtocounter{table}{1}

\section{Spectral energy distributions}
A Spectral Energy Distribution (SED) for BD+40\degr4124 was 
constructed by combining UV, optical, infrared, submm and radio 
data of BD+40\degr4124 from literature (Wesselius et al. 1982; 
Strom et al. 1972a; Terranegra et al. 1994; Hillenbrand et al. 
1992, 1995; Cohen 1972; Lorenzetti et al. 1983; Di Francesco 
et al. 1997; Corcoran \& Ray 1998; Skinner et al. 1993; Bertout 
\& Thum 1982) with our new ISO spectra (Fig.~6a). As can be seen 
from this plot, the optical to submm photometry forms a smooth 
curve. At radio wavelengths the slope of the energy distribution 
appears significantly flatter. We interpret this SED as the sum 
of three components: continuum emission from the central star in 
the UV and optical, thermal emission from dust heated by the 
central star in the infrared to submm and free-free emission 
from the stellar wind or an H\,{\sc ii} region at cm wavelengths. 
The SWS spectrum of BD+40\degr4124 shows a rising continuum 
component starting around 13~$\mu$m. This component is not 
present in the infrared photometry by Cohen (1972). 
This could be due to the smaller aperture size or the chopping 
throw used for background subtraction by Cohen. We interpret 
this second dust component, only visible in the ISO data, as 
diffuse emission throughout the star forming region.

There is general consensus in the literature that the spectral type 
of BD+40\degr4124 is about B2 Ve+sh (Merrill et al. 1932; Herbig 1960; 
Strom et al. 1972b; Swings 1981; Finkenzeller 1985; Hillenbrand et al. 
1995), corresponding to an effective temperature $T_{\rm eff}$ = 
22,000~K and surface gravity $\log g$ = 4.0 (Schmidt-Kaler 1982). 
With this spectral type and the optical photometry by Strom et al. 
(1972a) we arrive at a value of $E(B-V) = 0\fm97$ towards the optical 
star. This value agrees well with the $E(B-V)$ derived from the 
diffuse interstellar band at 5849\AA~(Oudmaijer et al. 1997). 
With this $E(B-V)$ value and assuming a normal interstellar 
extinction law (i.e. $R_V$ = $A_V/E(B-V)$ = 3.1), we corrected 
the data for extinction.

A Kurucz (1991) stellar atmosphere model with $T_{\rm eff}$ 
and $\log g$ corresponding to the star's spectral type 
(assuming solar abundance) was fitted to the 
extinction-corrected UV to optical SED. The fit is also 
shown in Fig.~6a. As can be seen from this figure, 
the SED fit is not perfect in the UV. Possibly a UV 
excess due to a strong stellar wind is present in 
BD+40\degr4124. The fitted 
Kurucz model was used to compute an apparent stellar 
luminosity $L/(4 \pi d^2)$ for BD+40\degr4124, which 
was converted to an absolute stellar luminosity using the 
BD+40\degr4124 distance estimate of 1~kpc by Shevchenko 
et al. (1991). Infrared luminosities were computed by 
fitting a spline to the infrared to mm data and subtracting 
the Kurucz model from this fit. 
The derived values for the optical and infrared luminosity 
of BD+40\degr4124 are $6.4 \times 10^3$ and 
$3.1 \times 10^2$~L$_\odot$, respectively. These luminosities 
place the star close to the zero-age main sequence position 
of a 10~M$_\odot$ star in the Hertzsprung-Russell diagram (HRD).

We followed the same procedure to construct a SED for 
LkH$\alpha$ 224, combining optical, infrared and 
submm photometry from literature (Strom et al. 1972a; 
Shevchenko, personal communication; Hillenbrand et al. 1995; 
Cohen 1972; Di Francesco et al. 1997) 
with our new ISO spectra. It is shown in Fig.~6b. Since 
LkH$\alpha$ 224 is strongly variable in the 
optical, the data used in the SED were selected to be 
obtained near maximum system brightness, when available. 
The optical to infrared photometry of LkH$\alpha$ 224 
forms a smooth curve in the SED, which we interpret as 
due to continuum emission from the star in the optical, 
and thermal emission from dust heated by the central star 
in the infrared.

There is no agreement in the literature on the spectral type of 
LkH$\alpha$ 224. Wenzel (1980) classified the star 
spectroscopically as B8e, whereas Hillenbrand et al. (1995) 
obtained a spectral type of B5 Ve. Corcoran \& Ray (1997) 
derived a spectral type of A0 for LkH$\alpha$ 224, based upon 
the relative strength of the He\,{\sc i} $\lambda$4471 and 
Mg\,{\sc ii} $\lambda$4481 lines. In the optical, LkH$\alpha$ 224 
shows large-amplitude photometric variations ($>$ 3\mag) due 
to variable amounts of circumstellar extinction (Wenzel 1980; 
Shevchenko et al. 1991, 1993). In the group of Herbig Ae/Be 
stars, these large-amplitude variations are only observed when 
the spectral type is A0 or later (van den Ancker et al. 1998a). 
Therefore we favour a later spectral type than B for LkH$\alpha$ 224. 
The detection of the Ca\,{\sc ii} H and K lines in absorption 
(Magakyan \& Movseyan 1997) also seems to favour a later 
spectral type than mid-B. Recent optical spectroscopy of 
LkH$\alpha$ 224 by de Winter (personal communication) 
shows strong, variable, P-Cygni type emission in H$\alpha$. 
The higher members of the Balmer series (up to H15) are 
present and in absorption, together with the Ca\,{\sc ii} H 
and K lines. Numerous shell lines are present in the 
spectrum as well, which could hinder spectral classification 
solely based upon red spectra. From the available optical 
spectroscopy, we classify LkH$\alpha$ 224 as A7e+sh, 
corresponding to $T_{\rm eff}$ = 7850~K. We estimate the 
uncertainty in this classification to be around 2 subclasses, 
or 400~K.
\begin{figure}[t]
\vspace*{0.5cm}
\centerline{\psfig{figure=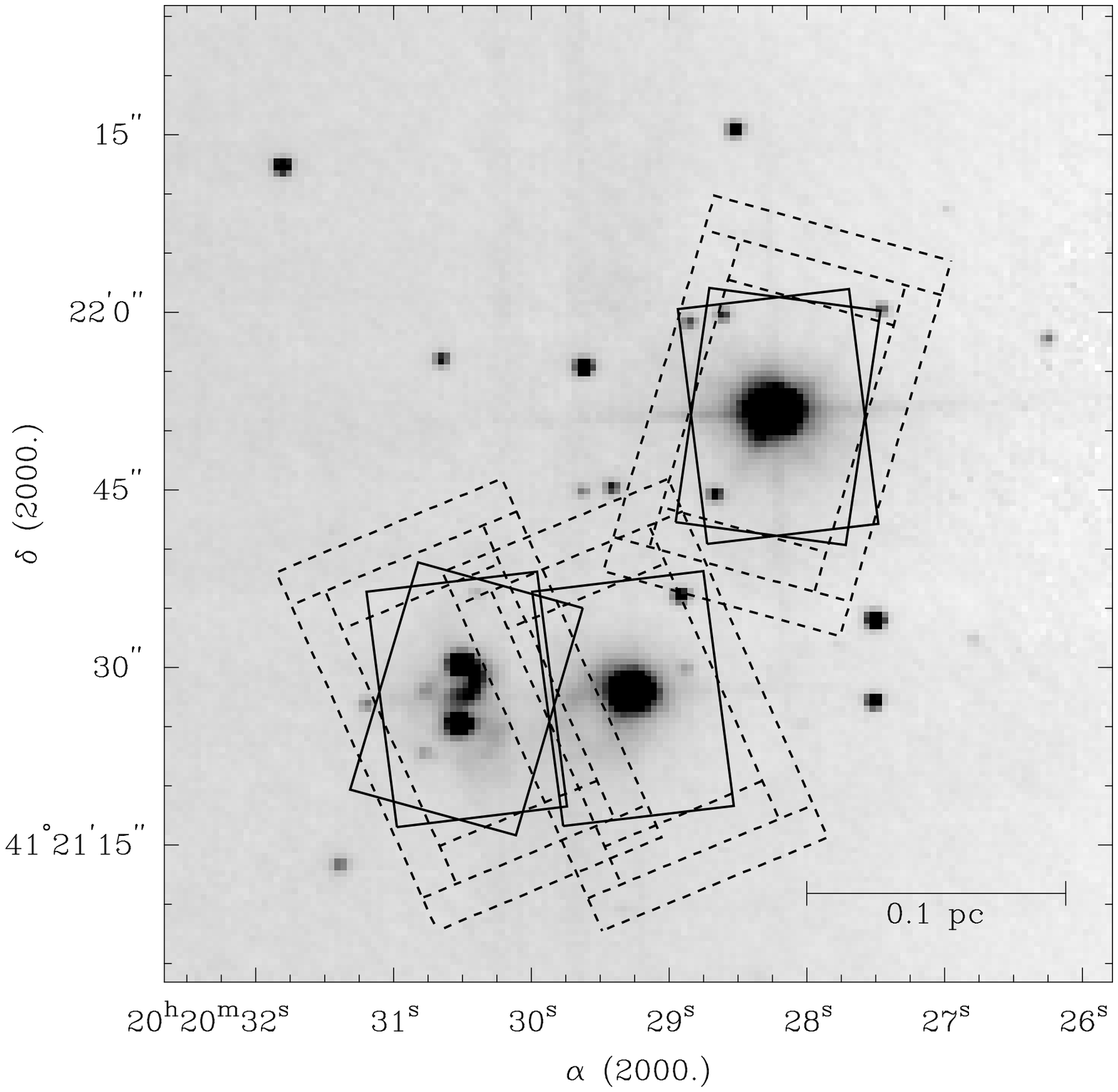,width=10.8cm,angle=270}}
\caption[]{SWS aperture positions for our measurements of (from 
right to left) BD+40\degr4124, LkH$\alpha$ 224 and LkH$\alpha$ 225 
superimposed on a K$^{\prime}$-band image of the region. The rectangles 
indicate the apertures (in increasing size) for SWS bands 1A--2C 
(2.4--12.0~$\mu$m), 3A--3D (12.0--27.5~$\mu$m), 3E (27.5--29.5~$\mu$m) 
and 4 (29.5--40.5~$\mu$m). Solid and dashed rectangles give the 
orientation of the SWS aperture for the AOT S02 and S01 observations, 
respectively. For clarity, only the smallest aperture is shown for 
the AOT S02 observations.}
\end{figure}
\begin{figure}[t]
\vspace*{0.5cm}
\centerline{\psfig{figure=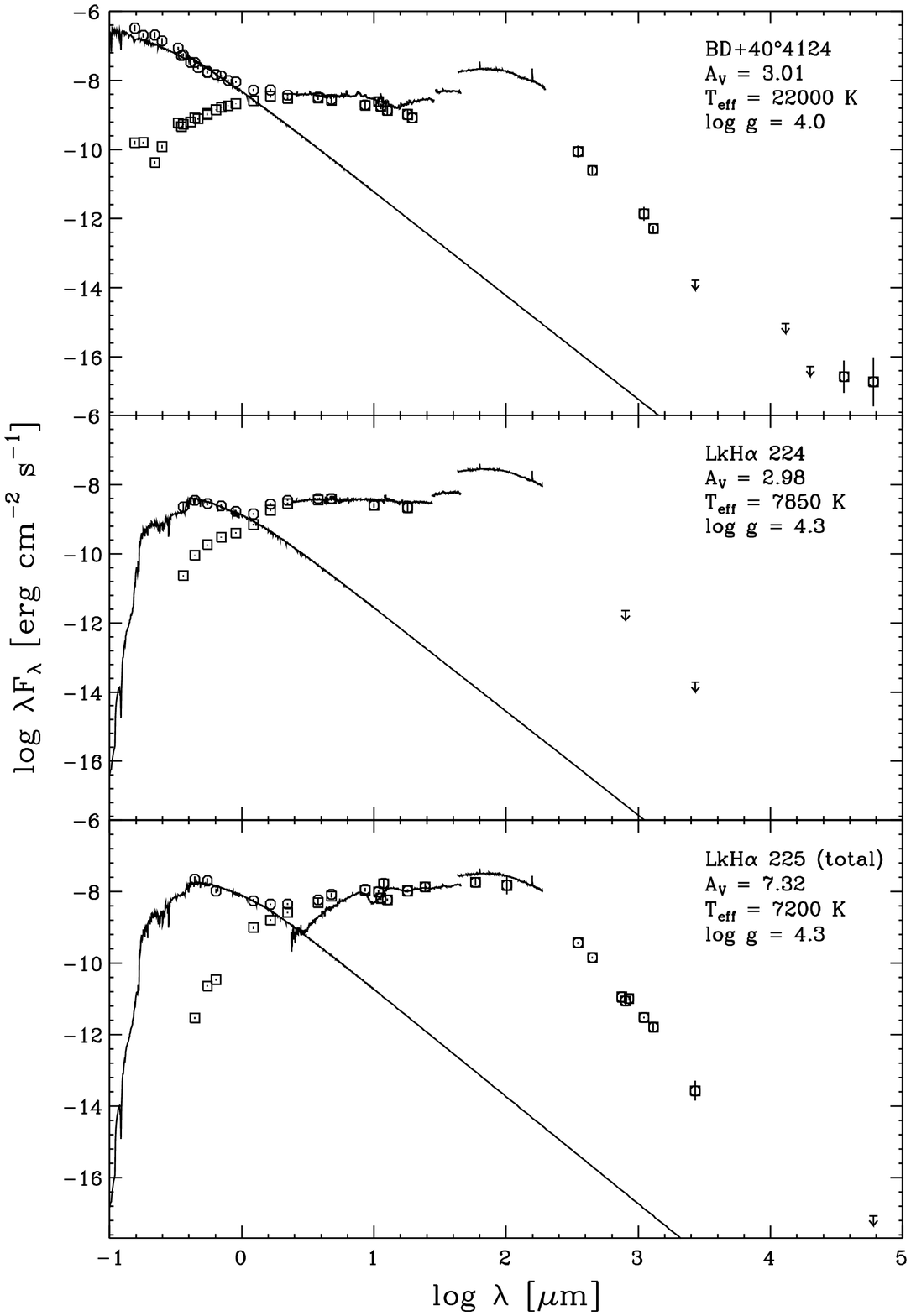,width=8.5cm,angle=0}}
\caption[]{Observed (squares) and extinction-corrected (circles) 
spectral energy distributions of BD+40\degr4124 (top), LkH$\alpha$ 224 
(middle) and LkH$\alpha$ 225 (bottom). Also shown (lines) are the 
ISO spectra and a Kurucz model fitted to the extinction-corrected 
UV to optical photometry for each object.}
\end{figure}

With the new spectral type of A7e+sh and the optical photometry 
by Strom et al. (1972a), we derive an $E(B-V)$ at maximum 
system brightness for LkH$\alpha$ 224 of 0\fm96. Again 
fitting a Kurucz model to the extinction-corrected photometry, 
we derive values for the optical and infrared luminosity 
of LkH$\alpha$ 224 of $1.1 \times 10^2$ and 
$2.7 \times 10^2$~L$_\odot$, respectively. The good 
fit of the SED to the Kurucz model with 
$T_{\rm eff}$ = 7850~K shows that the photometry is 
consistent with our adopted spectral type of A7e for 
LkH$\alpha$ 224. Comparing the derived parameters 
of LkH$\alpha$ 224 with the pre-main sequence evolutionary 
tracks by Palla \& Stahler (1993), we see that it 
is located in the HRD at the location of a 
$3 \times 10^5$~yr old 3.5~M$_\odot$ star. This brings 
the age of the star in good agreement with those 
of the low-mass members of the BD+40\degr4124 group 
(Hillenbrand et al. 1995).

Again the same procedure as above was followed to construct 
a SED for LkH$\alpha$ 225 (Fig.~6c), 
combining optical to radio measurements from literature 
(Shevchenko, personal communication; Hillenbrand et al. 1995; 
Cohen 1972; Weaver \& Jones 1992; Di Francesco et al. 1998; 
Henning et al. 1998; Cohen et al. 1982) with our ISO data. 
These data refer to the sum of all components in this object. 
Since LkH$\alpha$ 225 shows strong variability in the optical and 
infrared, only data obtained near maximum brightness was used 
in the SED. Like in BD+40\degr4124, the SED of LkH$\alpha$ 225 
consists of a smooth continuum ranging from the optical to the 
submm. In view of the knowledge of the nature of 
LkH$\alpha$ 225 from literature, this SED must be interpreted 
as the sum of the three components of LkH$\alpha$ 225 in 
the optical to submm, each consisting of a hot component 
with circumstellar dust emission. The plotted 
IRAS fluxes of IRAS 20187+4111 (Weaver \& Jones 1992) are 
smaller than those in the LkH$\alpha$ 225 LWS spectrum. Most 
likely, this is due to the background subtraction applied on 
the IRAS data. This means that a cool component is 
present as well throughout the region, covering an area 
that is larger than the size of the IRAS detectors. In 
view of submm maps of the region (Aspin et al. 
1994; Henning et al. 1998), this seems plausible.

As can be seen in Fig.~6c, below 8~$\mu$m the flux levels 
in the SWS spectrum are considerably smaller than those 
from ground-based photometry at maximum brightness. 
At longer wavelengths they agree. By extrapolating our ISO 
SWS spectrum, we deduce that the total $K$ band magnitude 
of LkH$\alpha$ 225 was about 8\fm3 at the time of our 
observation. This is within the range of values present 
in the literature. Above 10~$\mu$m, the SWS flux levels are 
in agreement with ground-based measurements, suggesting 
that the large-amplitude variability of LkH$\alpha$ 225 
is limited to the optical to near-infrared spectral range.

The spectral type of LkH$\alpha$ 225 is very uncertain. 
Hillenbrand et al. (1995) give a spectroscopic classification 
of mid A--Fe for both LkH$\alpha$ 225-North and 
LkH$\alpha$ 225-South, whereas Wenzel (1980) derives a 
spectral type of G--Ke for the sum of both components. 
Aspin et al. (1994) suggested the southern component to be 
of intermediate-mass, based on its high bolometric luminosity. 
Unlike for BD+40\degr4124 and LkH$\alpha$ 224, this 
makes the direct derivation of $E(B-V)$ very uncertain. 
Adopting $E(B-V)$ = 4\fm97, corresponding to 
$A_V = 15\fm4$ derived from the silicate feature in the 
next section, results in an optical slope in the energy 
distribution that is steeper than the Rayleigh-Jeans 
tail of a black body. This could be because the 
extinction at the time of the optical observations was 
smaller than during the ISO observations. Therefore we used 
a rather uncertain value of $E(B-V)$ = 2\fm36, based on 
the observed $(B-V)$ colour of LkH$\alpha$ 225 and the 
intrinsic colours of a F0 star, to correct the photometry 
of LkH$\alpha$ 225 for extinction.

To obtain a rough measure for the total luminosity of 
LkH$\alpha$ 225, we fitted a Kurucz model with 
$T_{\rm eff}$ = 7200~K, corresponding to a spectral type 
of F0, to the extinction-corrected optical photometry. 
This results in values of $5.9 \times 10^2$ and 
$1.6 \times 10^3$~L$_\odot$ for the total optical and 
infrared luminosities of LkH$\alpha$ 225. We conclude 
that at least one of the stars in the LkH$\alpha$ 225 
system must be a massive (5--7~M$_\odot$) 
pre-main sequence object in order to explain these 
luminosities.

\section{Solid-state features}
The ISO full grating scans of BD+40\degr4124 consist of a relatively 
smooth continuum, with a number of strong emission lines 
superimposed. The familiar unidentified infrared (UIR) emission 
features at 3.3, 6.2, 7.6, 7.8, 8.6 and 11.3~$\mu$m, 
often attributed to polycyclic aromatic hydrocarbons (PAHs), are 
prominently present. Possibly the 12.7~$\mu$m feature 
also observed in other sources showing strong PAH emission (Beintema 
et al. 1996) is present as well. No 9.7~$\mu$m amorphous silicate 
feature, either in emission or in absorption, is visible in the 
BD+40\degr4124 SWS spectrum. The slight curvature in the 
ground-based 8--13~$\mu$m BD+40\degr4124 spectrum obtained by 
Rodgers \& Wooden (1997), which they interpreted as a silicate 
feature in emission, is absent in our SWS data.

The continuum in BD+40\degr4124 seems to consist of two 
distinct components, one from 2.4--13~$\mu$m and the other 
from 13--200~$\mu$m. In the SWS range, it is well fit by 
the superposition of two blackbodies of 320 and 100~K. For the LWS 
spectrum a significant excess at the long-wavelength part remains 
after the black-body fit. Probably a range in temperatures is present 
starting from 100~K down to much lower temperatures. The jumps in the 
spectrum at the positions corresponding to a change in aperture suggest 
that this component is caused by diffuse emission in the star forming 
region.

Remarkably, the line flux measured in the 3.3~$\mu$m PAH feature 
of BD+40\degr4124 ($1.1 \times 10^{-14}$~W~m$^{-2}$) is much 
larger than the $2.0 \times 10^{-15}$~W~m$^{-2}$ upper limit 
obtained by Brooke et al. (1993). PAH features also appear 
absent in the 8--13~$\mu$m {\sc hifogs} spectrum shown by 
Rodgers \& Wooden (1997). This could either be an effect of the 
different aperture size used ($20\arcsec \times 14\arcsec$~for 
SWS versus a 1\farcs4 circular aperture by Brooke et al.) in case 
the PAH emission is spatially extended, or be due to 
time-variability of the PAH emission. If the first explanation is 
correct, the PAH emission should come from a region with a 
diameter larger than 3\farcs2 and hence be easily resolvable 
with ground-based imaging.

The ISO spectrum of LkH$\alpha$ 224 consists of one 
single smooth continuum, with a few emission lines superimposed. 
No solid-state emission or absorption is apparent in the SWS spectrum. 
Again the slight curvature seen in the {\sc hifogs} 8--13~$\mu$m 
spectrum of LkH$\alpha$ 224 (Rodgers \& Wooden 1997) is absent 
in our SWS data. In view of the fact that most Herbig Ae stars show 
strong silicate emission in the 10~$\mu$m range (e.g. van den Ancker 
1999), the absence of the silicate feature is quite remarkable. 
It means that the dust emission must come from a region which is 
completely optically thick up to infrared wavelengths and yet only 
produce 3 magnitudes of extinction towards the central star in the 
optical. It is obvious that such a region cannot be spherically 
symmetric, but must be highly flattened or have a hole through which 
we can seen the central 
star relatively unobscured. A dusty circumstellar disk might be 
the prime candidate for such a region. The absence of silicate 
emission in LkH$\alpha$ 224 might then be explained by either an 
intrinsically higher optical depth in its circumstellar disk 
than those surrounding other Herbig Ae stars, a view supported by 
its relative youth, or be caused by a nearly edge-on orientation 
of the circumstellar disk.

The infrared continuum of LkH$\alpha$ 224 cannot be fit by a 
single blackbody. Rather a sum of blackbodies with a range of 
temperatures starting at several hundreds Kelvins and continuing 
to temperatures of a few tens of K are required to fit the spectrum 
adequately. Again the jumps in the spectrum at the positions 
corresponding to a change in aperture are present, suggesting that 
at the longer wavelengths we are looking at diffuse emission in the 
star forming region. However, the slope of the continuum in 
LkH$\alpha$ 224 is significantly different from that of the cool 
component in BD+40\degr4124. This could either be due to a difference 
in the maximum temperature or to a difference in temperature gradient 
of the diffuse component between the two positions.

The SWS spectrum of LkH$\alpha$ 225 looks very different. 
Again we have a strong, smooth continuum with many emission lines, 
but now also absorption lines due to gaseous molecular material 
as well as strong absorption features due to solid-state 
material are present. The familiar 9.7~$\mu$m absorption feature 
due to the Si--O stretching mode in amorphous silicates is 
strong. Apart from a 20\% difference in the absolute flux 
levels, the silicate feature looks very similar to what was found 
by IRAS LRS (Olnon et al. 1986), suggesting that it is constant 
in time. The integrated optical depth ($\int \tau(\nu) d\nu$) 
of the 9.7~$\mu$m feature in the ISO SWS spectrum is 196~cm$^{-1}$. 
The 18~$\mu$m feature due to the bending mode 
of Si--O is also present in absorption with an integrated 
optical depth of 11~cm$^{-1}$. The O--H bending mode of water 
ice at 3.0 is present in absorption as well (156~cm$^{-1}$). 

Using intrinsic band strengths from literature (Tielens \& 
Allamandola 1987; Gerakines et al. 1995), we have converted the 
integrated optical depths of the 9.7~$\mu$m silicate and 
3.0~$\mu$m H$_2$O features to column densities. The results 
are a column of $1.6 \times 10^{18}$ molecules for the silicates 
and $7.8 \times 10^{17}$ molecules for H$_2$O ice. The 2:1 ratio 
of these two species is within the range of what is found 
in other lines of sight (Whittet et al. 1996 and references 
therein). Curiously, CO$_2$ ice, which invariably accompanies 
H$_2$O ice in all other lines of sight studied by ISO 
(Whittet et al. 1996; Boogert et al. 1999; Gerakines et al. 
1999) is absent in LkH$\alpha$ 225. The derived upper limit 
on the solid CO$_2$/H$_2$O ratio is 0.04, much less than 
the canonical value of 0.15.
\begin{figure*}[t]
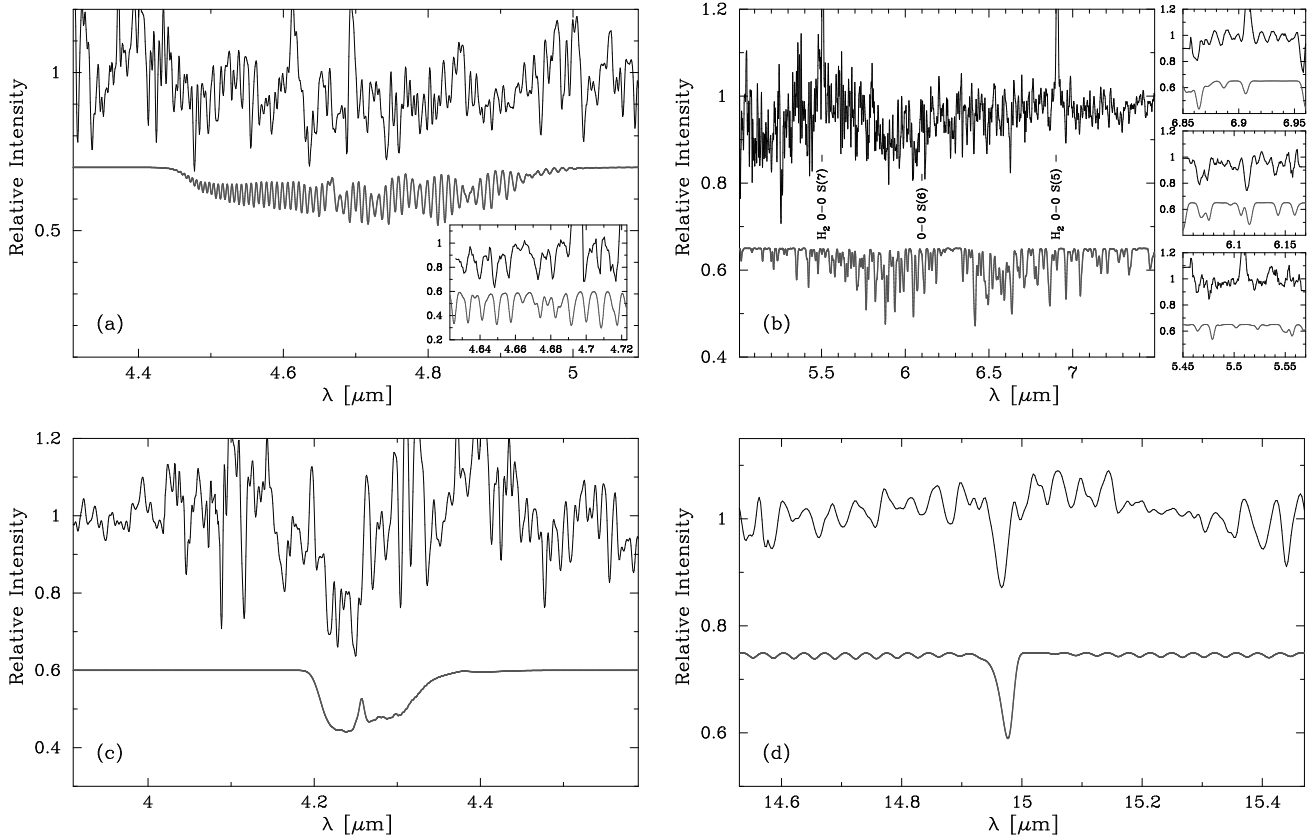

\centerline{\psfig{figure=m8961_f7a.ps,width=8.4cm,angle=270}
            \hspace*{0.3cm}
            \psfig{figure=m8961_f7b.ps,width=8.4cm,angle=270}}
\vspace*{0.3cm}
\centerline{\psfig{figure=m8961_f7c.ps,width=8.4cm,angle=270}
            \hspace*{0.3cm}
            \psfig{figure=m8961_f7d.ps,width=8.4cm,angle=270}}
\caption[]{Detected gas-phase molecular absorption bands in 
LkH$\alpha$ 225 with the continuum normalized to unity (top curves). 
Also shown (bottom curves) are the computed absorption spectra, shifted 
for clarity. The large plots show the AOT 01 SWS data, whereas the 
insets show the AOT 02 data containing molecular bands. $T$ and $b$ 
were required to be identical in all fits. 
(a) Observed CO ro-vibrational lines, compared to a $T_{\rm ex}$ = 300~K, 
$b$ = 5~km~s$^{-1}$, $N$(CO) = 10$^{20}$~cm$^{-2}$ model. 
(b) H$_2$O $\nu_2$ band, with a $T_{\rm ex}$ = 300~K, $b$ = 5~km~s$^{-1}$, 
$N$(H$_2$O) = 10$^{19}$~cm$^{-2}$ model. 
(c) gas-phase CO$_2$ $\nu_3$ band, with a $T_{\rm ex}$ = 300~K, 
$b$ = 5~km~s$^{-1}$, $N$(CO$_2$) = $3 \times 10^{17}$~cm$^{-2}$ model. 
(d) gas-phase CO$_2$ $\nu_2$ band, with a $T_{\rm ex}$ = 300~K, 
$b$ = 5~km~s$^{-1}$, $N$(CO$_2$) = $3 \times 10^{17}$~cm$^{-2}$ model.
}
\end{figure*}

No aperture jumps are visible in the SWS spectrum of LkH$\alpha$ 225. 
The bulk of the continuum flux must therefore come from 
an area that is small compared to the smallest SWS aperture 
($20\arcsec \times 14\arcsec$). The discontinuity between the 
SWS and LWS spectra shows that although LkH$\alpha$ 225 itself 
is probably the strongest far-infrared source in the region, 
the total far-infrared flux of the group is dominated by the 
diffuse component. Like in LkH$\alpha$ 224, the SWS spectrum 
of LkH$\alpha$ 225 cannot be fit well with a single blackbody.
Again a sum of blackbodies with a range of temperatures 
starting at several hundreds and continuing to 
temperatures of a few tens of K are required to fit the spectrum 
adequately.

PAHs appear absent in our LkH$\alpha$ 225 SWS spectrum. In view 
of the claim by Deutsch et al. (1994) that the southern component 
of LkH$\alpha$ 225 is the strongest UIR emitter in the region, 
this is remarkable. One possible explanation for this 
apparent discrepancy could be the large degree of variability 
($>$ 3\mag~in $K$; Allen 1973) of LkH$\alpha$ 225. 
In this case the component responsible for the PAH 
emission should have been much fainter than the other 
component at the time of observation. However, 
Aspin et al. (1994) showed the southern component to be 
dominant longward of 3~$\mu$m. Comparison of our 
LkH$\alpha$ 225 spectrum with infrared photometry by 
previous authors shows that the source was not near minimum 
brightness at the time of the SWS observation, making the 
explanation that the southern component of LkH$\alpha$ 225 was 
faint improbable. The most likely explanation for the apparent 
discrepancy between our non-detection and the reported UIR 
emission (Deutsch et al. 1994) might therefore be that the 
UIR emission within the southern component of LkH$\alpha$ 225 
is variable in time. Note that because the gas cooling 
timescale is much longer than the PAH cooling timescale, if 
PAH emission has been present within recent years, gas emission 
from a remnant PDR might be expected in the vicinity of 
LkH$\alpha$ 225.

Since extinction in the continuum surrounding the 9.7~$\mu$m feature 
is small compared to the extinction within this feature, the extinction 
$A_\lambda$ at wavelength $\lambda$ across a non-saturated 9.7~$\mu$m 
feature can simply be obtained from the relation 
$A_\lambda = -2.5 \log (I/I_0)$. Using an average interstellar extinction 
law which includes the silicate feature (Fluks et al. 1994), we 
can then convert these values of $A_\lambda$ to a visual extinction, 
resulting in a value of $A_V = 15\fm4 \pm 0\fm2$ toward LkH$\alpha$ 225. 
This value is significantly smaller than the $A_V \approx 50\mag$ 
obtained from the C$^{18}$O column density towards LkH$\alpha$ 225 
(Palla et al. 1995). Although smaller, our estimate of $A_V$ is still 
compatible with the value of $25\mag \pm 9\mag$ derived from the 
ratio of the 1--0 S(1) and Q(3) lines of H$_2$ (Aspin et al. 1994).

\section{Gas-phase molecular absorption lines}
In the SWS spectrum of LkH$\alpha$ 225 several absorption bands 
are present due to a large column of gas-phase CO, CO$_2$ and 
H$_2$O in the line of sight. They are shown in Fig.~7. Molecular 
absorption bands are absent in the BD+40\degr4124 and 
LkH$\alpha$ 224 spectra. We compared the observed gas-phase 
absorption lines to synthetic gas-phase absorption bands 
computed using molecular constants from the HITRAN 96 database 
(Rothmann et al. 1996). First the relative population of the 
different levels within one molecule were computed for an 
excitation temperature $T_{\rm ex}$. For the relative 
abundances of the isotopes, values from 
the HITRAN database were adopted. A Voigt profile was taken 
for the lines, with a Doppler parameter 
$b$ = $\frac{c}{2 \sqrt{\ln 2}}
\frac{\Delta \lambda}{\lambda}$ for the Gaussian component 
of the Voigt profile and the lifetime of the considered 
level times $b/4\pi$ for the Lorentz component. 
The optical depth $\tau_\lambda$ as a function of wavelength 
was then computed by evaluating the sum of all integrated 
absorption coefficients, distributed on the Voigt profile. 
Synthetic absorption spectra were obtained by computing 
$\exp(-\tau_\lambda)$. This procedure for computing 
absorption spectra is nearly identical to that followed 
by Helmich et al. (1996) and Dartois et al. (1998). 
For comparison with the observations, the spectra were 
convolved with the ISO SWS instrumental profile.
For the AOT 02 observations, the instrumental profile 
is nearly Gaussian with a wavelength-dependent resolving 
power $R = FWHM/\lambda$ ranging between 1400 and 3000 
(Leech et al. 1997). For our AOT 01 observations, the 
resolution is a factor four lower.

Best fits of the synthetic absorption spectra to the 
observed CO, CO$_2$ and H$_2$O bands are also shown in Fig.~7.
Because of the relatively low signal to noise in the spectra, 
the fits for the different absorption bands were not done 
independently, but $T_{\rm ex}$ and $b$ were required to 
be the same for all species. This procedure is only valid 
if the different molecular species are co-spatial. The results 
of similar fitting procedures in other lines of sight 
(e.g. van Dishoeck et al. 1996; Dartois et al. 1998; Boogert 
1999) suggest that this is the case. Satisfactory fits to the 
data could be obtained for excitation temperatures of 
a few 100~K and a Doppler parameter $b$ between 3 and 
10~km~s$^{-1}$. The derived column of warm CO$_2$ is 
several times 10$^{17}$~cm$^{-2}$. For H$_2$O it is 
about 10$^{19}$~cm$^{-2}$, whereas the column of gas-phase 
CO is in the range 10$^{19}$--10$^{21}$~cm$^{-2}$.

The derived values of $b$ are much higher than those 
expected from purely thermal broadening 
($0.1290 \sqrt{T_{\rm ex}/A}$~km~s$^{-1}$, with $A$ the 
molecular weight in amu; Spitzer 1978). This must be caused 
by macroscopic motion within the column of warm gas in our 
line of sight toward LkH$\alpha$ 225. The fact that our 
deduced values of $b$ are within the range of those found 
toward hot cloud cores (Helmich et al. 1996; van Dishoeck 
\& Helmich 1996; Dartois et al. 1998) suggests that it is 
due to turbulence within the cloud rather than a dispersion 
in large-scale motion such as infall or rotation.

\section{Hydrogen recombination lines}
\begin{figure}[t]
\centerline{\psfig{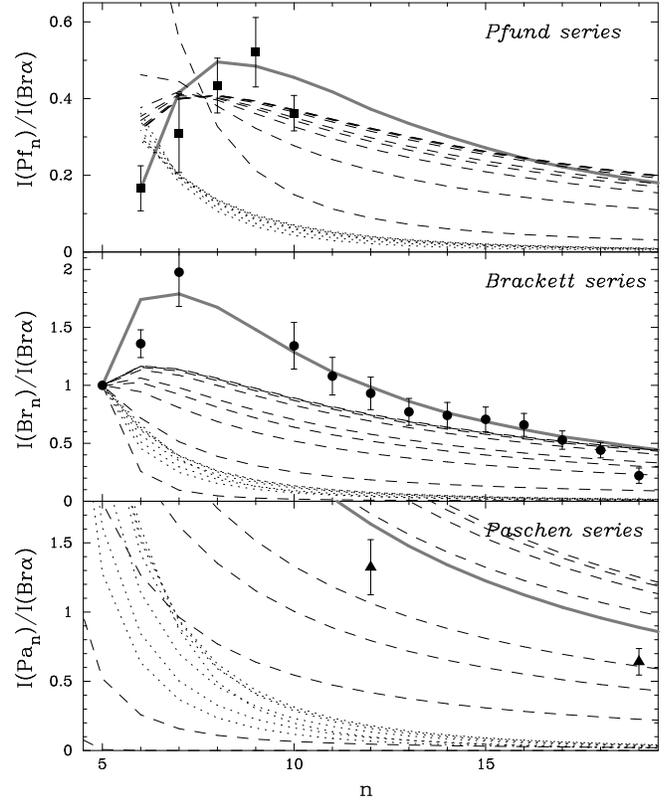}}
\caption[]{Extinction-corrected H\,{\sc i} line flux ratios in 
BD+40\degr4124 for (from top to bottom) the Pfund, Brackett and 
Paschen series. Also shown (dotted lines) are theoretical values 
for Case B recombination for $T$ = 1000--30,000~K and $n$ = 
10$^3$--10$^9$~cm$^{-3}$. The dashed lines give model 
predictions for an optically thick static slab of gas with 
temperatures (from bottom to top) between 1000 and 100,000~K. 
The gray solid line shows the best fit of the LTE fully ionized 
wind model.}
\end{figure}
The ISO SWS spectrum of BD+40\degr4124 contains several emission 
lines of the Pfund and Brackett series of H\,{\sc i}. No H\,{\sc i} 
lines were detected in LkH$\alpha$ 224 and LkH$\alpha$ 225. The 
lines fluxes for BD+40\degr4124 were corrected for extinction 
using the value of $A_V$ = 3\fm0 derived in section 3. We 
combined these data with ground-based measurements of emission lines 
in the Brackett and Paschen series (Harvey 1984; Hamann \& Persson 
1992; Rodgers \& Wooden 1992) for BD+40\degr4124 to create plots 
of the H\,{\sc i} line flux ratio as a function of upper level 
quantum number $n$ (Fig.~8). The fact that the values for the 
Brackett series lie on a continuous curve shows that they are 
not influenced much by the different beam sizes used, indicating 
that they must originate in a compact region.

In Fig.~8 we also show predicted values for Menzel \& Baker 
(1938) Case B recombination for a range of temperatures from 
1000 to 30,000~K and densities from 10$^3$ to 10$^9$~cm$^{-3}$, 
taken from Storey \& Hummer (1995). The fit of the Case B 
values to the data is poor, indicating that at least the 
lower lines in the Pfund and Brackett series are not optically 
thin, as assumed in Case B recombination theory, but have 
$\tau$ $\ga$ 1. An error in our estimate of the extinction 
towards the line emitting region cannot reconcile our observations 
with the Case B values. The poor fit of the Case B model 
predictions to the observed H\,{\sc i} line fluxes means 
that the lines do not originate in the low-density H\,{\sc ii} 
region responsible for the [O\,{\sc iii}] emission (section 8).

In the limit that all lines are optically thick, the line 
flux ratios of a static slab of gas in local thermodynamic 
equilibrium (LTE) will be dominated by the wings of the line 
profile. Such a slab of gas will produce line fluxes 
$I_\lambda$ $\propto$ $B_\lambda(T_e)$ $f_{ij}^{2/5}$, with 
$B_\lambda$ the Planck function for an electron temperature 
$T_e$, and $f_{ij}$ the oscillator strength for the transition 
(Drake \& Ulrich 1980). In Fig.~8 we also show these values 
for a range of electron temperatures between 1000 and 100,000~K. 
Again this simple model cannot reproduce the observed line flux 
ratios in BD+40\degr4124, especially for Pf$\alpha$/Br$\alpha$ and 
Br$\gamma$/Br$\alpha$. The most likely explanation for the 
poor fit is that the lines are more broadened than due to the 
Stark wings of the line profile, probably due to macroscopic 
motion of the gas.

The poor fit of both the Case B values and the optically thick 
static slab model shows that we are not looking at an optically 
thick static slab of gas, but at an optically thick ionized 
gas with a significant velocity dispersion. Most 
likely it originates either in an ionized wind, or in an 
ionized gaseous disk surrounding BD+40\degr4124. Without 
information on the kinematics of the gas, the distinction 
between these two cases is hard to make. To qualitatively 
show that such a model can indeed reproduce the observations, 
we also show in Fig.~8 the best fit of a simple LTE model for 
a totally ionized wind following the treatment of 
Nisini et al. (1995). This best fit model has a rather 
high mass loss rate ($\approx$ 10$^{-6}$~M$_\odot$~yr$^{-1}$), 
arising from a compact region ($\approx$ 12 R$_\star$) in 
radius. However, the assumption of LTE will cause this 
simple model to significantly overestimate the mass loss rate 
(Bouret \& Catala 1998), whereas the assumption that the 
wind is completely ionized will cause an underestimate of 
the size of the line emitting region. Therefore the values 
given here should be approached as lower, respectively 
upper limits. Also, a model for an ionized gaseous disk 
surrounding BD+40\degr4124 may give equally good fits 
to the data.

If the lower lines in the Pfund and Brackett series are 
optically thick, the region in which these lines originate 
should also produce optically thick free-free emission 
at radio wavelengths. In the simple wind model employed 
above, the Br$\alpha$ line flux is related to the 6~cm 
continuum flux by the relation $S_\nu$ (mJy) = 
$1.4 \times 10^{14}$ $F_{Br\alpha}$ (W~m$^{-2}$) 
$v_2^{-1/3}$, with $v_2$ the wind velocity in units 
of 100~km~s$^{-1}$ (Simon et al. 1983). Assuming 
$v_2 = 1$, this relation yields 0.39~mJy for 
BD+40\degr4124, in excellent agreement with the observed
6~cm flux of 0.38~mJy (Skinner et al. 1993). Although 
highly simplified, this analysis does show that the region 
responsible for the H\,{\sc i} line emission can also 
produce sufficient amounts of radio continuum emission to 
account for the observed fluxes and that our interpretation 
of the radio component in the SED as free-free emission 
(section 3) is correct.

\section{Molecular hydrogen emission}
At all three observed positions in the BD+40\degr4124 region 
we detected emission from pure rotational lines of molecular 
hydrogen. They are listed in Table~1 and are shown in Figs.~2--4. 
For BD+40\degr4124 and LkH$\alpha$ 224, 0--0 S(0) to S(5) were 
detected. For LkH$\alpha$ 225 pure rotational lines up to S(10) 
as well as several ro-vibrational lines were detected. 
It is clear that in LkH$\alpha$ 225 the 0--0 S(6) line 
escaped detection because of its unfortunate location in 
the H$_2$O gas phase $\nu_2$ absorption band (Fig.~7b). The 
line flux of the 0--0 S(5) line may also be slightly 
($\approx$ 10\%) underestimated because of this. 
We have detected significantly more lines than the preliminary 
results reported in the A\&A ISO First Results issue 
(Wesselius et al. 1996), both because of improvements in the 
SWS data reduction since 1996 and because additional data were 
obtained. Therefore a new analysis of these H$_2$ lines is in place. 

From the H$_2$ line fluxes $I(J)$ it is possible to calculate 
the apparent column densities of molecular hydrogen in the 
upper $J$ levels, averaged over the SWS beam, $N(J)$, using 
$N(J) = \frac{4\pi I(J)}{A_{ij}} \frac{\lambda}{hc}$, with $\lambda$ 
the wavelength, $h$ Planck's constant and $c$ the speed of light. 
The transition probabilities $A_{ij}$ were taken from Turner et al. 
(1977). Line fluxes were corrected for extinction using 
$I(J) = I_{\rm obs} 10^{A(\lambda)/2.5}$ using the average 
interstellar extinction law by Fluks et al. (1994). For 
both BD+40\degr4124 and LkH$\alpha$ 224 we used 
$A_V$ = 3\fm0, valid for the optical stars. For 
LkH$\alpha$ 225 we used the $A_V$ = 15\fm4 derived in 
section 4.
\begin{figure}[t]
\centerline{\psfig{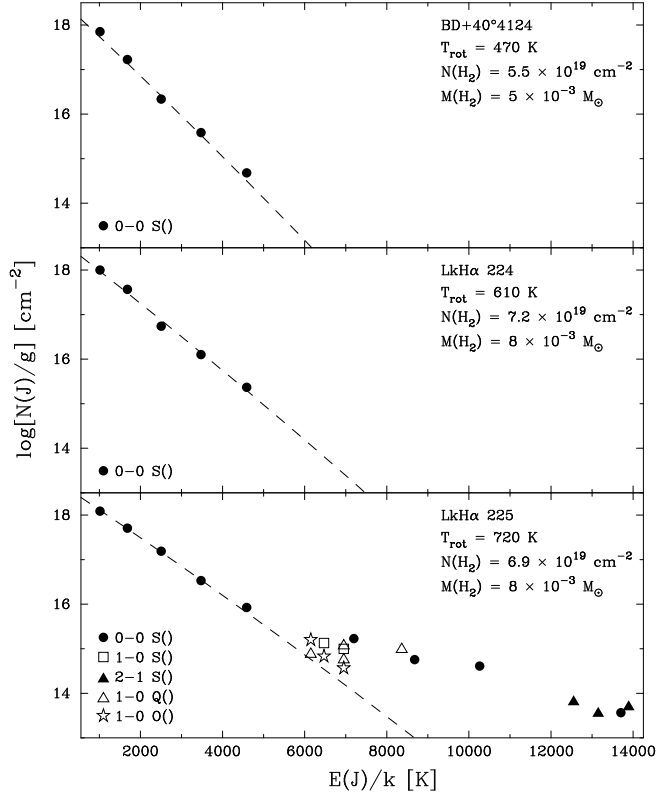}}
\caption[]{H$_2$ excitation diagrams for BD+40\degr4124 (top), 
LkH$\alpha$ 224 (middle) and LkH$\alpha$ 225 (bottom). ISO observations 
of pure rotational lines are indicated by the filled dots. Squares, 
triangles and stars indicate measurements of ro-vibrational lines from 
the SWS spectra and from Aspin et al. (1994). Formal errors for most 
measurements are smaller than the plot symbol. The dashed lines gives 
the Boltzmann distribution fits to the low-lying pure rotational lines.}
\end{figure}

A useful representation of the H$_2$ data is to plot the log of
$N(J)/g$, the apparent column density in a given energy 
level divided by its statistical weight, versus the energy of the 
upper level. The statistical weight $g$ 
is the combination of the rotational and nuclear spin components. 
It is $(2J+1)$ for para-H$_2$ (odd $J$) and $(2J+1)$ times the 
Ortho/Para ratio for ortho-H$_2$ (even $J$). We have assumed the 
high temperature equilibrium relative abundances of 3:1 for the 
ortho and para forms of H$_2$ (Burton et al. 1992). 
The resulting excitation diagrams are shown in Fig.~9. For 
LkH$\alpha$ 225 we also included the measurements of the 
H$_2$ ro-vibrational lines by Aspin et al. (1994). 
For the lines in common, their values differ somewhat with 
ours, although the line ratios agree. This is probably an effect 
of their smaller beam size and indicates that in our measurements 
the beam filling factor is smaller than one. In Fig.~9 the values 
of Aspin et al. were scaled to our measurements.
\begin{figure}[t]
\centerline{\psfig{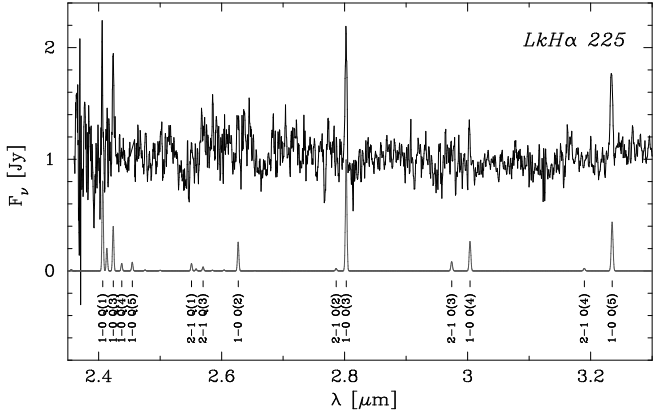}}
\caption[]{Continuum-subtracted 2.35--3.30~$\mu$m SWS spectrum of 
LkH$\alpha$ 225. Also shown (bottom curve) is a  
synthetic H$_2$ spectrum with $T_{\rm exc}$ = 710~K, corrected 
for an extinction of $A_V = 15\fm4$.} 
\end{figure}

In case the population of the energy levels is thermal, all 
apparent column densities should lie on a nearly straight line 
in the excitation diagram. The slope of this line is inversely 
proportional to the excitation temperature, while the intercept is 
a measure of the total column density of warm gas. As can be seen 
from Fig.~9, this is indeed the case in all three stars for the 
pure rotational lines with an upper level energy up to 5000~K. 
The fact that the points for ortho and para H$_2$ lie on the 
same line proves that our assumption on their relative abundances 
is correct. We have used the formula for the H$_2$ column density 
for a Boltzmann distribution given by Parmar et al. (1991) to fit 
our data points in the low-lying pure rotational levels, using the 
rotational constants given by Dabrowski (1984), varying the excitation 
temperature and column density. The results are shown as the dashed 
lines in Fig.~9.

Van den Ancker et al. (1998b and in preparation) have employed 
predictions of H$_2$ emission from PDR, J-shock and C-shock models 
(Burton et al. 1992; Hollenbach \& McKee 1989; Kaufman \& Neufeld 
1996), to determine the excitation temperature from the low-lying pure 
rotational levels as a function of density $n$ and 
either incident FUV flux $G$ (in units of the average interstellar 
FUV field G$_0$ = $1.2 \times 10^{-4}$ erg~cm$^{-2}$~s$^{-1}$~sr$^{-1}$; 
Habing 1968) or shock velocity $v_s$ in an identical way as was done 
for the observations presented here. They arrived at the conclusion 
that the PDR and J-shock models allow a fairly small (200--540~K) 
range of excitation temperatures, whereas for C-shocks 
this range is much larger (100--1500~K). In the model predictions 
for shocks, $T_{\rm exc}$ does not depend much on density, 
whereas for PDRs it does not depend much on $G$. Once the mechanism of 
the H$_2$ emission is established, it can therefore be used to 
constrain $v_s$ or $n$ in a straightforward way.

The values for $T_{\rm exc}$ measured for 
LkH$\alpha$ 224 and LkH$\alpha$ 225 fall outside the range of 
$T_{\rm exc}$ values predicted by either PDR or J-shock models. 
They are similar to the values found in other regions containing 
shocks (van den Ancker et al. 1998b). If the observed pure 
rotational H$_2$ emission is indeed due to a planar C-shock, its 
shock velocity should be around 20~km~s$^{-1}$. In section 4 
we saw that our BD+40\degr4124 SWS full grating scan contains strong 
emission in the PAH bands. This indicates that a PDR is 
present at that location. If we assume that the observed H$_2$ 
emission is entirely due to this PDR, the observed $T_{\rm exc}$ of 
470~K points to a rather high ($\ga 5 \times 10^5$) density within 
the PDR.

For LkH$\alpha$ 225, ro-vibrational as well as pure rotational 
emission from H$_2$ was detected (Fig.~10). Both the higher pure 
rotational lines and the ro-vibrational lines significantly 
deviate from the straight line defined by the lower pure rotational 
lines in the excitation diagram and do not form a smooth line by 
themselves, as expected in 
the case of multiple thermal components. Such a situation can 
either occur because of fluorescence of H$_2$ in a strong UV 
field (Black \& van Dishoeck 1987; Draine \& Bertoldi 1996), 
be due to X-ray or EUV heating (Wolfire \& K\"onigl 1991; 
Tin\'e et al. 1997), or be due to the formation energy of 
re-formed H$_2$ in post-shock gas (Shull \& Hollenbach 1978).
Although the H$_2$ spectra resulting from the possible formation 
mechanisms do differ in the relative prominence of transitions 
with $\Delta v \ge 2$ (Black \& van Dishoeck 1987), the 
distinction between these situations is rather difficult 
to make with the detected lines in LkH$\alpha$ 225. 

By summing all the apparent column densities for the 
ro-vibrational and higher pure rotational lines we obtain 
an estimate for the observed total column of non-thermal 
H$_2$ in LkH$\alpha$ 225 of $2 \times 10^{16}$~cm$^{-2}$, 
corresponding to $2 \times 10^{-6}$~M$_\odot$ or 0.3 earth masses 
within the SWS beam. This value of $N$(H$_2^*$) is within the 
range predicted by UV fluorescence models. However, the 
absence of PAH emission in LkH$\alpha$ 225 at the time 
of the ISO observations suggests that there was no strong UV 
radiation field at that time. Therefore we consider the 
possibility that the ro-vibrational H$_2$ emission 
originates in a shock more likely.

Interestingly, in LkH$\alpha$ 225 we have detected two pairs of 
lines arising from the same energy level: the 1--0 Q(1) \& 1--0 O(3) 
and the 1--0 Q(3) \& 1--0 O(5) lines. For these pairs of lines 
$N(J)$ must be identical. Therefore they can be used to derive a 
differential extinction between the two wavelengths of the lines 
using $A(\lambda_1)-A(\lambda_2) = 
2.5 \log\left(\frac{I_2 \lambda_2 A_{ij,1}}{I_1 \lambda_1 A_{ij,2}}\right)$.
Using the interstellar extinction curve by Fluks et al. 
(1994), we can then convert the differential extinction to a value 
of $A_V$. The results of this procedure are values for $A_V$ of 
$60\mag \pm 20\mag$ and $40\mag \pm 20\mag$ for the 
1--0 Q(1) \& 1--0 O(3) and 1--0 Q(3) \& 1--0 O(5) line pairs, 
respectively. These extinction values are somewhat 
larger than the $15\fm4 \pm 0\fm2$ derived from the silicate feature 
in section 4. But they agree well with $A_V \approx 50\mag$ 
obtained from the C$^{18}$O column density towards LkH$\alpha$ 225 
(Palla et al. 1995). If the slope of the infrared extinction law 
is approximately interstellar, the extinction towards the region 
producing the low-lying pure rotational 
H$_2$ lines in LkH$\alpha$ 225 must be smaller than this value 
though; such a large extinction would cause the extinction-corrected 
0--0 S(3) line to deviate significantly from the thermal 
distribution line in the excitation diagram, since it is located 
within the 9.7~$\mu$m silicate feature in the extinction curve.
We conclude that in LkH$\alpha$ 225 either there must be two distinct 
sources of H$_2$ emission, a 710~K thermal component suffering 
little extinction -- presumably a C-shock -- and a heavily 
extincted, deeply embedded component -- possibly a J-shock, or 
the slope of the infrared extinction law towards this region must 
be much steeper than interstellar.

\section{Carbon-monoxide emission lines}
\begin{figure}[t]
\centerline{\psfig{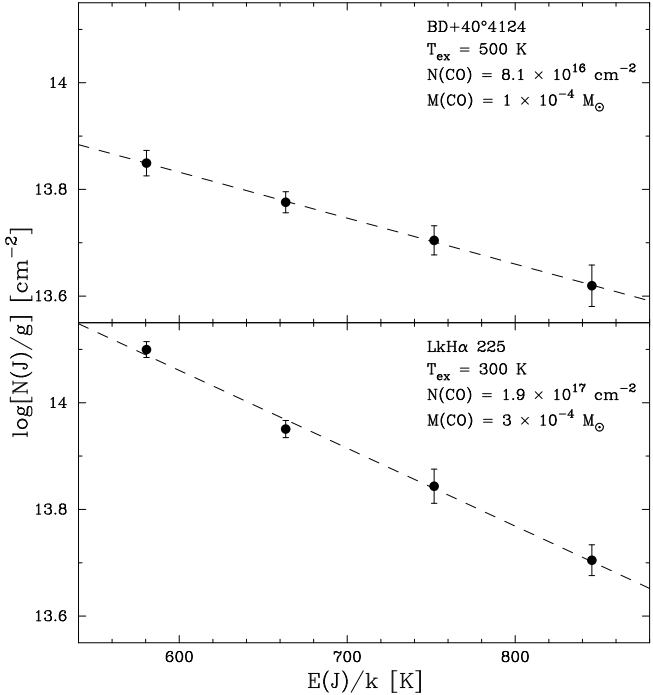}}
\caption[]{CO excitation diagrams for BD+40\degr4124 (top) and 
LkH$\alpha$ 225 (bottom). The dashed lines give the fit of the 
Boltzmann distribution to the data points.} 
\end{figure}
At both the positions of BD+40\degr4124 and LkH$\alpha$ 225, 
several emission lines due to the $v$ = 0--0 transitions of 
gas-phase CO were detected in the long-wavelength part of the 
LWS spectra (Table~2). Similar to what was done for the H$_2$ 
emission in the previous section, we constructed CO excitation 
diagrams, using molecular data from Kirby-Docken \& Liu (1978). 
They are shown in Fig.~11. The temperature and column of CO 
resulting from the Boltzmann fit to these excitation diagrams 
are 500~K and $8.1 \times 10^{16}$~cm$^{-2}$ and 300~K and 
$1.9 \times 10^{17}$~cm$^{-2}$ for BD+40\degr4124 and 
LkH$\alpha$ 225, respectively. We estimate the errors in these 
fit parameters to be around 50~K in temperature and 50\% in 
CO column.

The observed CO lines have critical densities of around 
10$^6$~cm$^{-3}$. Therefore the observation of these lines 
also implies densities of at least this order of magnitude 
in the originating region. The CO excitation temperature of 500~K 
found at the position of BD+40\degr4124 is comparable to that 
found from the H$_2$ lines. The CO/H$_2$ mass fraction of 0.01 is 
compatible for what one would expect for the warm gas in a PDR. 
However, assuming that densities of around 10$^6$~cm$^{-3}$ would 
exist in the entire BD+40\degr4124 PDR would be implausible. We 
therefore conclude that if these CO lines and the H$_2$ both 
originate in the large-scale environment of BD+40\degr4124, 
the PDR must have a clumpy structure (e.g. Burton et al. 1990).

The CO emission arising from the neighbourhood of LkH$\alpha$ 225 
is remarkably different from that seen in H$_2$: here the excitation 
temperature of the CO is much lower than that found in H$_2$. 
The CO/H$_2$ mass fraction of 0.03 is also much higher than that 
found at the position of BD+40\degr4124. It seems 
likely that another component than the C-shock responsible for 
the H$_2$ emission needs to be invoked to explain the low 
CO excitation temperature. Again this region must have 
densities of around 10$^6$~cm$^{-3}$. A possible disk 
or extended envelope around LkH$\alpha$ 225 could have both 
the temperature and densities required to explain the observed 
CO spectrum.

\section{Atomic fine structure lines}
\begin{figure}[t]
\centerline{\psfig{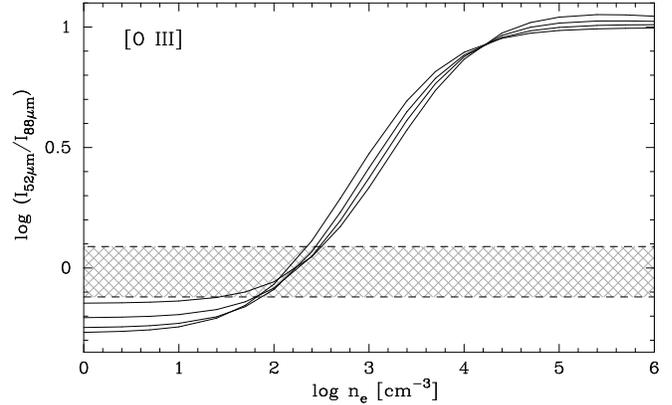}}
\caption[]{Behaviour of the [O\,{\sc iii}] 51.8~$\mu$m/88.4~$\mu$m 
line flux ratio as a function of electron density $n_e$ (solid lines). 
The curves for temperatures between 5,000 and 50,000~K (from right to 
left) are nearly identical. The hatched region between the dashed lines 
shows the error interval of the observed [O\,{\sc iii}] ratio in 
BD+40\degr4124.} 
\end{figure}
In order to interpret our observations as arising in either a PDR, 
C-Shock or J-shock, we compare our results with theoretical models of 
such regions (Tielens \& Hollenbach 1985; Hollenbach \& McKee 1989; 
Kaufman \& Neufeld 1996). Important constraints come from the observed 
fine structure lines. The mere presence of certain lines helps us 
to identify the mechanism responsible for the observed emission. 
In contrast to both C- and J-shocks, PDRs do not produce significant 
quantities of [S\,{\sc i}] emission (Tielens \& Hollenbach 1985). 
C-shocks only contain trace fractions of ions and hence cannot explain 
the [Fe\,{\sc ii}], [Si\,{\sc ii}] or [C\,{\sc ii}] emission. 
\begin{table*}
\caption[]{Model parameters for the emission line fits.}
\tabcolsep0.17cm
\begin{flushleft}
\begin{tabular}{lccccccccccc}
\hline\noalign{\smallskip}
Object          & \multicolumn{3}{c}{PDR} && \multicolumn{3}{c}{C-shock} && \multicolumn{3}{c}{J-shock}\\
\noalign{\vspace{0.02cm}}
\cline{2-4}\cline{6-8}\cline{10-12}\noalign{\vspace{0.04cm}}
                & $n$ [cm$^{-3}$] & $G$ [G$_0$] & $\Omega$ [10$^{-8}$ sr] &  
                & $n$ [cm$^{-3}$] & $v_s$ [km~s$^{-1}$] & $\Omega$ [10$^{-8}$ sr] & 
                & $n$ [cm$^{-3}$] & $v_s$ [km~s$^{-1}$] & $\Omega$ [10$^{-8}$ sr]\\
\noalign{\smallskip}
\hline\noalign{\smallskip}
BD+40\degr4124  & 10$^4$/10$^6$ & 10$^4$/10$^5$ & 0.55/0.05 && -- & -- & -- && -- & -- & --\\
LkH$\alpha$ 224 & -- & -- & -- && $\approx$ 10$^5$ & 20 & $\approx$ 0.10 && -- & -- & --\\
LkH$\alpha$ 225 & -- & -- & -- && $\approx$ 10$^5$ & 20 & $\approx$ 0.20 && 10$^4$--10$^5$ & 60 & 0.12\\
\noalign{\smallskip}
\hline
\end{tabular}
\end{flushleft}
\end{table*}

In BD+40\degr4124 the H$_2$ lines and the presence of PAH emission 
strongly suggest the presence of a PDR. No [S\,{\sc i}] was detected, 
so no shock is required to explain the emission from this region. 
The [Fe\,{\sc ii}] and [Si\,{\sc ii}] emission observed at the 
position of BD+40\degr4124 may also originate in a PDR. 
The [O\,{\sc iii}] emission lines detected by SWS cannot be 
produced in a PDR. The ratio of the [O\,{\sc iii}] 52 and 88~$\mu$m 
lines is sensitive to the electron density $n_e$ in the line forming 
region. It hardly depends on the temperature. We computed the expected 
ratio for a range of temperatures and densities, in the approximation 
that we can treat [O\,{\sc iii}] as a five level ion following 
the treatment by Aller (1984), using collisional strengths from 
Aggarwal et al. (1982). The results are shown in Fig.~12. 
We conclude that the [O\,{\sc iii}] lines observed by LWS are 
formed in a H\,{\sc ii} region with 
$n_e$ = 0.6--2.2 $\times$ $10^2$~cm$^{-3}$. In view of the 22,000~K 
effective temperature of BD+40\degr4124, it is surprising that 
[O\,{\sc iii}] emission is detected, since simple photo-ionization 
models predict that for stars with $T_{\rm eff}$ $<$ 32,000~K most 
oxygen should remain neutral. Clearly another ionization model must be 
present here. Possibly the region also responsible for the H\,{\sc i} 
lines could collisionally ionize the oxygen and we are observing the 
outer, low-density part of an ionized wind.

Apart from the PDR, a H\,{\sc ii} region surrounding BD+40\degr4124 may also 
contribute significantly to the observed [Fe\,{\sc ii}] and [Si\,{\sc ii}] 
emission. Therefore we have used the photo-ionization code {\sc cloudy} 
(version 90.04; Ferland 1996) to generate model predictions for line strengths 
for an H\,{\sc ii} region surrounding BD+40\degr4124. A Kurucz (1991) model 
with parameters appropriate for BD+40\degr4124 was taken as the input 
spectrum, and a spherical geometry and a constant electron density 
of 2 $\times$ $10^2$~cm$^{-3}$ throughout the H\,{\sc ii} region 
were assumed. According to this model, 30\% of the observed 
[Si\,{\sc ii}] 34.82~$\mu$m flux and 40\% of the 25.99~$\mu$m [Fe\,{\sc ii}] 
flux will originate in the H\,{\sc ii} region. From comparing the remainder 
of the fluxes of the atomic fine structure lines to the PDR models of 
Tielens \& Hollenbach (1985), we derive that the PDR should be fairly 
dense ($\approx$ 10$^4$~cm$^{-3}$) and have an incident far-UV field 
of about 10$^4$~G$_0$. The extent of this PDR should be around 180 square 
arcseconds, or about half of the SWS beam.

In view of the observed H$_2$ spectrum (section 7), a C-shock appears the 
most likely candidate for explaining the observed [S\,{\sc i}] emission 
in LkH$\alpha$ 224. A C-shock with a shock velocity of around 
20~km~s$^{-1}$ can explain both the H$_2$ and [S\,{\sc i}] emission. 
However, the density and extent of the region are poorly constrained: 
the emission may either arise in a dense ($\approx$ 10$^6$~cm$^{-3}$) 
region of only a few square arcseconds, or arise in a region of 
about 10$^3$--10$^4$~cm$^{-3}$ with a high beam filling factor. Since 
it is observed in the largest SWS aperture, the observed [Si\,{\sc ii}] 
emission could be due to contamination from LkH$\alpha$ 225 (see Fig.~5).

For LkH$\alpha$ 225 the situation appears more complicated. The detection 
of [S\,{\sc i}] clearly shows that a shock must be present in the 
region. A C-shock can explain the observed thermal H$_2$ component, 
with similar parameters as for LkH$\alpha$ 224, but can only explain part 
of the [S\,{\sc i}] emission. The [Fe\,{\sc ii}] and [Si\,{\sc ii}] 
emission could either come from a deeply embedded 
H\,{\sc ii} region, a deeply embedded PDR, or a J-shock. In view of 
the necessity to have an additional source of [S\,{\sc i}] emission, the 
most likely candidate might be a J-shock. A J-shock model with $v_s$ 
$\approx$ 60~km~s$^{-1}$ and $n$ $\approx$ 10$^4$--10$^5$~cm$^{-3}$ can 
indeed reproduce the observed [S\,{\sc i}] and [Si\,{\sc ii}] intensities. 
However, to reproduce the observed [Fe\,{\sc ii}] 26.0~$\mu$m strength, 
iron would have to be about eight times more depleted than assumed 
in the Tielens \& Hollenbach PDR models. The observed [C\,{\sc ii}] 
157.7~$\mu$m emission at the position of LkH$\alpha$ 225 cannot be 
explained by shocks. This could very well be due to contamination 
by the BD+40\degr4124 PDR or the galactic [C\,{\sc ii}] background, 
however.

Fit parameters for the PDR, J-shock and C-shock models used to 
explain the H$_2$, CO and fine structure emission lines are listed 
in Table~3. The summed model predictions for the line fluxes are 
listed in Table~2.

\section{Discussion and conclusions}
In the previous sections we have seen that in a small OB 
association like the BD+40\degr4124 group, different 
phenomena contribute to at first glance similar line spectra. 
The emission at the position of BD+40\degr4124 is well 
reproduced by the combination of a H\,{\sc ii} region around 
BD+40\degr4124 with a PDR. In Section 7 we derived values of the 
far-UV radiation field for its PDR from the observed emission 
lines. It is useful to compare those values to those expected 
from the central star to see whether that is a sufficient source 
of radiation, or another exciting source needs to be invoked. 
From the Kurucz (1991) model with $T_{\rm eff}$ = 22,000~K and 
$\log g = 4.0$, fitted to the extinction-corrected UV to optical 
SED of BD+40\degr4124 (section 3), 
we compute that BD+40\degr4124 has a total far-UV (6--13.6 eV) 
luminosity of $3.2 \times 10^3$~L$_\odot$. At a location 
$5 \times 10^3$~AU from the central star, this field will have 
diluted to the value of $10^5$~G$_0$ obtained from the infrared 
lines. At the BD+40\degr4124 distance of 1~kpc, this 
corresponds to an angular separation of 5\arcsec, consistent 
with the constraints imposed by the SWS aperture. We conclude 
that BD+40\degr4124 can indeed be responsible for the incident 
UV flux upon the PDR observed in that vicinity. Compared to 
the far-UV luminosity of BD+40\degr4124, the other two sources 
only show little UV emission (3 and 6~L$_\odot$ for 
LkH$\alpha$ 224 and 225, respectively), explaining why we 
don't observe a PDR at those positions.
\begin{table*}
\caption[]{Comparison of gas-phase and solid-state absorption column densities (in cm$^{-2}$) towards YSOs.}
\tabcolsep0.13cm
\begin{flushleft}
\begin{tabular}{lccccccccccccc}
\hline\noalign{\smallskip}
Object          & \multicolumn{3}{c}{gas-phase column} &\,& \multicolumn{3}{c}{solid column} &\,& \multicolumn{3}{c}{gas/solid ratio} &\,& Ref.\\
\noalign{\vspace{0.02cm}}
\cline{2-4}\cline{6-8}\cline{10-12}\noalign{\vspace{0.04cm}}
                & CO  & CO$_2$  & H$_2$O  && CO  & CO$_2$  & H$_2$O  && CO  & CO$_2$  & H$_2$O  && \\  
\noalign{\smallskip}
\hline\noalign{\smallskip}
LkH$\alpha$ 225 & 10$^{19}$--10$^{21}$       & 1--5 $\times$ 10$^{17}$    & 0.5--3 $\times$ 10$^{19}$ &
                & $<$ 6 $\times$ 10$^{17}$   & $<$ 3 $\times$ 10$^{16}$   & 7.8 $\times$ 10$^{17}$    &
                & $>$ 16                     & $>$ 3                      & $\approx$ 13              && (1)\\

T Tau           & 10$^{18}$--10$^{19}$       & $<$ 1 $\times$ 10$^{17}$   & $<$ 1 $\times$ 10$^{18}$  &
                & $<$ 5 $\times$ 10$^{17}$   & 1.4 $\times$ 10$^{17}$     & 5.4 $\times$ 10$^{17}$    &
                & $>$ 20                     & $<$ 0.7                    & $<$ 1.9                   && (2)\\
AFGL 4176       & 2 $\times$ 10$^{19}$       & 5 $\times$ 10$^{15}$       & 2 $\times$ 10$^{18}$      &
                & $<$ 5 $\times$ 10$^{16}$   & 1.2 $\times$ 10$^{17}$     & 9 $\times$ 10$^{17}$      &
                & $>$ 400                    & $\approx$ 0.04             & $\approx$ 2.2             && (3,4,5,6)\\
W33A            & $4.0 \times 10^{19}$       &  --                        & $1.0 \times 10^{18}$      &
                & $9.0 \times 10^{17}$       & $1.5 \times 10^{18}$       & $2.8 \times 10^{19}$      & 
                & $\approx$ 40               &  --                        & $\approx$ 0.04            && (3,7,8,9)\\
AFGL 2136       & 1.0 $\times$ 10$^{19}$     & 1 $\times$ 10$^{16}$       & 2 $\times$ 10$^{18}$      &
                & 1.8 $\times$ 10$^{17}$     & 6.1 $\times$ 10$^{17}$     & 5.0 $\times$ 10$^{18}$    &
                & $\approx$ 50               & $\approx$ 0.02             & $\approx$ 0.4             && (3,4,5,6)\\
RAFGL 7009S     & 6.1 $\times$ 10$^{18}$     & 1.0 $\times$ 10$^{17}$     & $\ge$ 2 $\times$ 10$^{18}$&
                & 1.8 $\times$ 10$^{18}$     & 2.5 $\times$ 10$^{18}$     & 1.1 $\times$ 10$^{19}$    &
                & $\approx$ 3.4              & $\approx$ 0.04             & $\ge$ 0.18                && (10)\\
Elias 29        & $> 1 \times 10^{19}$       &  --                        & $> 9 \times 10^{17}$      &
                & $1.7 \times 10^{17}$       & $6.5 \times 10^{17}$       & $3.0 \times 10^{18}$      & 
                & $>$ 55                     &  --                        & $>$ 0.3                   && (9)\\
AFGL 2591       & 1.1 $\times$ 10$^{19}$     & 1 $\times$ 10$^{16}$       & 2 $\times$ 10$^{18}$      &
                & $<$ 4 $\times$ 10$^{16}$   & 2.7 $\times$ 10$^{17}$     & 1.7 $\times$ 10$^{18}$    &
                & $>$ 270                    & $\approx$ 0.04             & $\approx$ 1.1             && (3,4,5,6)\\
NGC 7538 IRS9   & 1.4 $\times$ 10$^{19}$     & 8.0 $\times$ 10$^{15}$     & $<$ 3 $\times$ 10$^{17}$  &
                & 1.0 $\times$ 10$^{18}$     & 1.2 $\times$ 10$^{18}$     & 8.0 $\times$ 10$^{18}$    &
                & $\approx$ 14               & $\approx$ 0.01             & $<$ 0.04                  && (3,4,5,6)\\
\noalign{\smallskip}
\hline
\end{tabular}
\end{flushleft}
\vspace*{-0.2cm}
\noindent
References: 
(1) This paper; 
(2) van den Ancker et al. (1999); 
(2) Mitchell et al. (1990); 
(3) van Dishoeck et al. (1996); 
(4) de Graauw et al. (1996b); 
(5) van Dishoeck \& Helmich (1996); 
(6) Chiar et al. (1998); 
(7) Gerakines et al. (1999); 
(8) Boogert (1999); 
(9) Dartois et al. (1998).
\end{table*}

For LkH$\alpha$ 224 we showed that the infrared emission-line spectrum 
could be explained by a single non-dissociative shock. A C-shock 
with very similar parameters may also be present at the position 
of LkH$\alpha$ 225. Comparing our ISO SWS aperture positions 
(Fig.~5) with the image of the bipolar outflow centered on 
LkH$\alpha$ 225 (Palla et al. 1995), we note that this outflow 
covers both LkH$\alpha$ 224 and LkH$\alpha$ 225. One 
possible identification for the C-shock in both LkH$\alpha$ 224 
and LkH$\alpha$ 225 might therefore be the shock created as the 
outflow originating from LkH$\alpha$ 225 drives into the surrounding 
molecular cloud.

Apart from the large-scale C-shock, we have seen that probably 
a J-shock is present in LkH$\alpha$ 225 as well. This can be 
naturally identified with the Herbig-Haro knot LkH$\alpha$ 225-M 
observed in [S\,{\sc ii}] in the optical (Magakyan \& Movseyan 1997). 
It remains unclear whether the ro-vibrational H$_2$ lines
observed in this object are due to H$_2$ fluorescence by 
Ly$\alpha$ photons or to H$_2$ re-formation in the post-shock gas. 
Spectroscopy of lines with higher $v$ is required to clarify this.

In the line of sight toward LkH$\alpha$ 225, a large column consisting 
of warm gas-phase CO, CO$_2$ and H$_2$O and solid water ice and 
silicates was detected. The column of gas-phase water was found to 
be about 10 times higher that that of water ice. No CO$_2$ ice was 
detected toward LkH$\alpha$ 225, demonstrating that the bulk of 
the CO$_2$ is in the gas-phase. In Table~4 we compare 
the derived gas/solid ratio for LkH$\alpha$ 225 with those 
found in the lines of sight towards other young stellar objects. 
Both the H$_2$O and CO$_2$ gas/solid ratios are much higher than 
those found in other YSOs. For CO$_2$, the lower limit to the 
gas/solid is even a factor 100 higher than that found in the 
source with the previously reported highest ratio! 
At this point we can only speculate as to why this is the 
case. One possible explanation could be that the radiation field 
of BD+40\degr4124 is heating the outer layers of the LkH$\alpha$ 225 
envelope, evaporating the CO$_2$ ice. If this is the case, 
BD+40\degr4124 should be located closer to the observer than 
LkH$\alpha$ 225, suggesting that LkH$\alpha$ 225 is located at 
the far side of the ``blister'' in which the BD+40\degr4124 group 
is located. Another explanation could be 
that the LkH$\alpha$ 225 core itself is unusually warm. Since the 
abundance of gas-phase water is sensitive to the temperature 
(Charnley 1997), this would also explain the high abundance of 
water vapour in the line of sight toward LkH$\alpha$ 225. Because 
LkH$\alpha$ 225 is less massive than the luminous objects in 
which the previous gas/solid ratios have been determined, this 
high temperature might reflect a more evolved nature of the 
LkH$\alpha$ 225 core. Future 
research, such as a temperature determination from gas-phase 
CO measurements, should be able to distinguish between these 
possibilities and show whether we have discovered the chemically 
most evolved hot core known to date, or are looking through a 
line of sight particularly contaminated by a nearby OB star.

\acknowledgements{
The authors would like to thank Ewine van Dishoeck for her help in 
obtaining the BD+40\degr4124 LWS data and Dolf de Winter for making 
his optical spectra of LkH$\alpha$ 224 available to us. We thank 
Adwin Boogert and Issei Yamamura for helpful discussions on the 
molecular absorption spectra. Lynne Hillenbrand kindly provided 
the K$^{\prime}$-band image of the region shown in Fig.~5. An 
anonymous referee provided useful suggestions for improvements 
of the paper. MvdA acknowledges financial support from NWO grant 
614.41.003 and through a NWO {\em Pionier} grant to L.B.F.M. Waters. 
This research has made use of the Simbad data base, operated at CDS, 
Strasbourg, France.}


\begin{thebibliography}{}
\bibitem{}
Aggarwal, K.M., Baluja, K.L., Tully, J.A. 1982, MNRAS 201, 923
\bibitem{}
Allen, D.A. 1973, MNRAS 161, 145
\bibitem{}
Aller, L.H. 1984, {\it ``Physics of Thermal Gaseous Nebulae''}, 
 Dordrecht, Reidel
\bibitem{}
Aspin, C., Sandell, G., Weintraub, D.A. 1994, A\&A 282, L25
\bibitem{}
Beintema, D.A., van den Ancker, M.E., Molster, F.J. et al.
 1996, A\&A 315, L369
\bibitem{}
Bertout, C., Thum, C. 1982, A\&A 107, 368
\bibitem{}
Black, F.H., van Dishoeck. E.F. 1987, ApJ 322, 412
\bibitem{}
Boogert, A.C.A. 1999, PhD thesis, Groningen University
\bibitem{}
Boogert, A.C.A., Ehrenfreund, P., Gerakines, P. et al. 1999, A\&A, in press
\bibitem{}
Bouret, J.C., Catala, C. 1998, A\&A 340, 163
\bibitem{}
Brooke, T.Y., Tokunaga, A.T., Strom, S.E. 1993, AJ 106, 656
\bibitem{}
Burton, M.G., Hollenbach, D.J., Tielens, A.G.G.M. 1990, ApJ 365, 620
\bibitem{}
Burton, M.G., Hollenbach, D.J., Tielens, A.G.G.M. 1992, ApJ 399, 563
\bibitem{}
Charnley, S.B. 1997, ApJ 481, 396
\bibitem{}
Chiar, J.E., Gerakines, P.A., Whittet, D.C.B. et al. 1998, ApJ 498, 716
\bibitem{}
Cohen, M. 1972, ApJ 173, L61
\bibitem{}
Cohen, M., Bieging, J.H., Schwartz, P.R. 1982, ApJ 253, 707
\bibitem{}
Corcoran, M., Ray, T.P. 1996, A\&A 321, 189
\bibitem{}
Corcoran, M., Ray, T.P. 1998, A\&A 331, 147
\bibitem{}
Dabrowski, I. 1984, Canadian J. Phys. 62, 1639
\bibitem{}
Dartois, E., d'Hendecourt, L., Boulanger, F., Jourdain de Muizon, M., 
 Breitfellner, M., Puget, J.L., Habing, H.J. 1998, A\&A 331, 651
\bibitem{}
de Graauw, Th., Haser, L.N., Beintema, D.A. et al. 1996a, A\&A 315, L49
\bibitem{}
de Graauw, Th., Whittet, D.C.B., Gerakines, P.A. et al.
 1996b, A\&A 315, L345
\bibitem{}
Deutsch, L.K., Iyengar, M.A., Hora, J.L., Hoffmann, W.F., Dayal, A., 
 Butner, H.M., Fazio, G.G. 1994, AAS 185, 8416
\bibitem{}
Di Francesco, J., Evans, N.J., Harvey, P.M., Mundy, L.G., Guilloteau, S., 
 Chandler, C.J. 1997, ApJ 482, 433
\bibitem{}
Draine, B.T., Bertoldi, F. 1996, ApJ 468, 269
\bibitem{}
Drake, S.A., Ulrich, R.K. 1980, ApJS 42, 351
\bibitem{}
Ferland, G.J. 1996, Univ. of Kentucky Physics Department Internal Report
\bibitem{}
Finkenzeller, U. 1985, A\&A 151, 340
\bibitem{}
Fluks, M.A., Plez, B., Th\'e, P.S., de Winter, D., Westerlund, B.E., 
 Steenman, H.C. 1994, A\&AS 105, 311
\bibitem{}
Gerakines, P.A., Schutte, W.A., Greenberg, J.M., van Dishoeck, E.F. 
 1995, A\&A 296, 810
\bibitem{}
Gerakines, P.A., Whittet, D.C.B., Ehrenfreund, P. et al. 1999, ApJ 522, 357
\bibitem{}
Habing, H.J. 1968, Bull. Astron. Inst. Netherlands 19, 421
\bibitem{}
Hamann, F., Persson, S.E. 1992, ApJS 82, 285
\bibitem{}
Harvey, P.M. 1984, PASP 96, 297
\bibitem{}
Helmich, F.P., van Dishoeck, E.F., Black, J.H. et al.
 1996, A\&A 315, L173
\bibitem{}
Henning, T., Burkert, A., Launhardt, R., Leinert, C., Stecklum, B. 
 1998, A\&A 336, 565
\bibitem{}
Herbig, G.H. 1960, ApJS 4, 337
\bibitem{}
Hillenbrand, L.A., Strom, S.E., Vrba, F.J., Keene, J. 1992, ApJ 397, 613
\bibitem{}
Hillenbrand, L.A., Meyer, M.R., Strom, S.E., Skrutskie, M.F. 1995, 
 AJ 109, 280
\bibitem{}
Hollenbach, D.J., McKee, C.F. 1989, ApJ 342, 306
\bibitem{}
Kaufman, M.J., Neufeld, D.A. 1996, ApJ 456, 611
\bibitem{}
Kessler, M.F., Steinz, J.A., Anderegg, M.E. et al.
 1996, A\&A 315, L27
\bibitem{}
Kirby-Docken, K., Liu, B. 1978, ApJS 36, 359
\bibitem{}
Kurucz, R.L. 1991, in {\it ``Stellar atmospheres--Beyond classical models''}
 (eds. A.G. Davis Philip, A.R. Upgren, K.A. Janes), L. Davis press, 
 Schenectady, New York, p. 441
\bibitem{}
Leech, K. et al. 1997, ``SWS Instrument Data Users Manual'', Issue 3.1, 
 SAI/95-221/Dc
\bibitem{}
Lorenzetti, D., Saraceno, P., Strafella, F. 1983, ApJ 264, 554
\bibitem{}
Magakyan, T.Y., Movsesyan, T.A. 1997, Pis'ma Astron. Zh. 23, 764 
 (Astron. Lett. 23, 666)
\bibitem{}
Menzel, D.H., Baker, J.G. 1938, ApJ 88, 52
\bibitem{}
Merrill, P.W., Humason, M.L., Burwell, C.G. 1932, ApJ 76, 156
\bibitem{}
Mitchell, G.F., Maillard, J.P., Allen, M., Beer, R., Belcourt, K. 1990, ApJ 363, 554
\bibitem{}
Nisini, B., Milillo, A., Saraceno, P., Vitali, F. 1997, A\&A 302, 169
\bibitem{}
Olnon, F.M., Raimond, E. \& IRAS Science Team 1986, A\&AS 65, 607
\bibitem{}
Oudmaijer, R.D., Busfield, G., Drew, J.E. 1997, MNRAS 291, 797
\bibitem{}
Palla, F., Stahler, S.W. 1993, ApJ 418, 414
\bibitem{}
Palla, F., Testi, L., Hunter, T.R., Taylor, G.B., Prusti, T., Felli, M., 
 Natta, A., Stanga, R.M. 1995, A\&A 293, 521
\bibitem{}
Parmar, P.S., Lacy, J.H., Achtermann, J.M. 1991, ApJ 372, L25
\bibitem{}
Rodgers, B., Wooden, D.H. 1997, AAS 191, 4709
\bibitem{}
Rothmann, L.S., Gamache, R.R., Tipping, R.H. et al. 1996, 
 J. Quant. Spectr. Radiat. Transfer 48, 469
\bibitem{}
Schmidt-Kaler, Th. 1982, in ``Landolt B\"ornstein Catalogue'', Vol VI/2b   
\bibitem{}
Shevchenko, V.S., Ibragimov, M.A., Chernysheva, T.L. 1991, Astron. Zh. 68, 
 466 (SvA 35, 229)
\bibitem{}
Shevchenko, V.S., Grankin, K.N., Ibragimov, M.A., Melnikov, S.Y., 
 Yakubov, S.D. 1993, Ap\&SS 202, 121
\bibitem{}
Shull, J.M., Hollenbach, D.J. 1978, ApJ 220, 525
\bibitem{}
Simon, M., Felli, M., Cassar, L., Fischer, J., Massi, M. 1983, ApJ 266, 623
\bibitem{}
Skinner, S.L., Brown, A., Stewart, R.T. 1993, ApJS 87, 217
\bibitem{}
Spitzer, L. 1978, ``Physical Processes in the Interstellar Medium'', 
 Wiley Interscience, New York
\bibitem{}
Storey, P.J., Hummer, D.G. 1995, MNRAS 272, 41
\bibitem{}
Strom, K.M., Strom, S.E., Breger, M., Brooke, A.L., Yost, J., 
 Grasdalen, G., Carrasco, L. 1972a, ApJ 173, L65
\bibitem{}
Strom, S.E., Strom, K.M., Yost, J., Carrasco, L., Grasdalen, G. 
 1972b, ApJ 173, 353
\bibitem{}
Swings, J.P. 1981, A\&AS 43, 331
\bibitem{}
Terranegra, L., Chavarr\'{\i}a, C., Diaz, S., Gonzalez-Patino, D. 
 1994, A\&AS 104, 557
\bibitem{}
Tielens, A.G.G.M., Allamandola, L.J. 1987, in {\it ``Interstellar Processes''}, 
 eds. D.J. Hollenbach \& H.A. Thronson Jr. (Dordrecht: Reidel), p. 397
\bibitem{}
Tielens, A.G.G.M., Hollenbach, D.J. 1985, ApJ 291, 722
\bibitem{}
Tin\'e, S., Lepp, S., Gredel, R., Dalgarno, A. 1997, ApJ 481, 282
\bibitem{}
Trams, N. et al. 1997, ``ISO-LWS Instrument Data Users Manual'', Issue 5.0, 
 SAI/95-219/Dc
\bibitem{}
Turner, J., Kirby-Docken, K., Dalgarno, A. 1977, ApJS 35, 281
\bibitem{}
van den Ancker, M.E., de Winter, D., Tjin A Djie, H.R.E. 1998a, 
 A\&A 330, 145
\bibitem{}
van den Ancker, M.E., Wesselius, P.R., Tielens, A.G.G.M., Waters, 
 L.B.F.M. 1998b, in {\it ``ISO's View on Stellar Evolution''}, 
 eds. L.B.F.M. Waters et al., Ap\&SS 255, 69
\bibitem{}
van den Ancker, M.E. 1999, PhD thesis, University of Amsterdam
\bibitem{}
van den Ancker, M.E., Wesselius, P.R., Tielens, A.G.G.M., van Dishoeck, 
 E.F., Spinoglio, L. 1999, A\&A 348, 877
\bibitem{}
van Dishoeck, E.F., Helmich, F.P. 1996, A\&A 315, L177
\bibitem{}
van Dishoeck, E.F., Helmich, F.P., de Graauw, Th. et al.
 1996, A\&A 315, L349
\bibitem{}
Weaver, W.B., Jones, G. 1992, ApJS 78, 239
\bibitem{}
Wenzel, W. 1980, Mitt. Ver. Sterne 8, 182
\bibitem{}
Wesselius, P.R., van den Ancker, M.E., Young, E.T. et al.
 1996, A\&A 315, L197
\bibitem{}
Wesselius, P.R., van Duinen, R.J., de Jonge, A.R.W., Aalders, J.W.G., 
 Luinge, W., Wildeman, K.J. 1982, A\&AS 49, 427
\bibitem{}
Whittet, D.C.B., Schutte, W.A., Tielens, A.G.G.M. et al.
 1996, A\&A 315, L357
\bibitem{}
Wolfire, M.G., K\"onigl, A. 1991, ApJ 383, 205
\end{thebibliography}
\end{document}